\documentclass[11pt]{article}
\usepackage[a4paper,margin=2.5cm]{geometry}
\usepackage[T1]{fontenc}
\usepackage[utf8]{inputenc}
\usepackage{lmodern}
\usepackage{microtype}
\usepackage{amsmath,amssymb}
\usepackage{booktabs}
\usepackage{threeparttable}
\usepackage{array}
\usepackage{graphicx}
\usepackage[round,authoryear]{natbib}
\usepackage{xurl}
\usepackage[hidelinks]{hyperref}
\usepackage{orcidlink}
\newcommand{\pkg}[1]{\textsf{#1}}
\newcommand{\proglang}[1]{\textsf{#1}}
\newcommand{\code}[1]{\texttt{#1}}
\newcommand{\sectionref}[1]{Section~\ref{#1}}
\newcommand{\tableref}[1]{Table~\ref{#1}}
\newcommand{\figureref}[1]{Figure~\ref{#1}}
\title{Foundation or Formula? A Simulation-Based Comparison of Google's TimesFM-3 and Classical Time-Series Models}
\author{Ebrahim Khaled Ebrahim\,\orcidlink{0009-0006-7839-8778} \and
Somaia Mohamed Ali\,\orcidlink{0009-0002-0940-8451} \and
Ahmed El-Kotory\,\orcidlink{0009-0000-1485-0768}\\[4pt]
\normalsize Department of Statistics, Faculty of Business, Alexandria University, Alexandria, Egypt\\
\normalsize Correspondence: \href{mailto:ebrahimkhaled@alexu.edu.eg}{ebrahimkhaled@alexu.edu.eg}}
\date{}
\begin{document}
\maketitle
\begin{abstract}
Time-series foundation models such as Google's TimesFM-3 forecast series they were never
trained on. Whether they should replace classical methods such as ARIMA and exponential smoothing is
difficult to settle on public benchmarks, because few benchmark datasets are absent from every
model's pre-training corpus. This exploratory study describes, under controlled conditions, how
TimesFM-3 behaves relative to classical methods and in which regimes it may be preferred. The
evaluation data are generated from nine known processes, so the model cannot have seen the series.
Across 7\,200 series and a pre-registered protocol with a 12-step horizon, neither family dominates.
TimesFM-3's error is never more than 1.31 times that of the best method in a scenario, whereas every classical
method's is at least 2.2 times somewhere; the largest classical failures occur with only two seasonal
cycles, where the automatic methods fall back to non-seasonal models, and without those scenarios automatic
ARIMA's worst case is 1.62. With four or more cycles automatic ARIMA is 21--24\% more accurate than
TimesFM-3 on seasonal ARIMA data. Among five foundation models, this
robustness is specific to TimesFM-3, and it is specific to short horizons: at steps 25 to 48, the three foundation models run at that horizon forecast a decline on a saturating process
whose level stays flat, and automatic ARIMA has the smaller worst case elsewhere. On intermittent demand TimesFM-3 shows no advantage over
simple benchmarks built from the history. In post-hoc checks, randomised parameters, heavy-tailed noise and
outliers, alone or together, leave these patterns qualitatively unchanged, although testing for
seasonality narrows the classical worst case; on the official test period of
1\,000 M4 monthly series TimesFM-3 is level with the fifth- to seventh-ranked M4 entries, behind the four
best, and on 101 macroeconomic series observed after the models' documented training data it is
statistically level with the classical methods. Because TimesFM-3 is a black box, the study characterises
its behaviour, not the reasons for it; the findings are conditional on the processes and horizons examined,
and the model's weights are licensed for non-commercial use only.
\end{abstract}

\section{Introduction}
\label{sec:intro}

Time-series foundation models are large neural networks, pre-trained on many series and then applied
to new series without task-specific training. On 31 August 2026 Google Research released TimesFM-3, a
330-million-parameter decoder-only transformer and the first member of its family designed for
multivariate forecasting, which its developers report as the top-ranked pre-trained model on three
public benchmarks \citep{timesfm3blog}; such claims are hard to verify, for the reason given below. For
applied forecasting the relevant question is whether such a model should be used, in preference to a
classical method such as ARIMA, for the series at hand; the two families also differ in cost,
interpretability, the basis of their prediction intervals and the terms under which they may be used.

Published benchmarks are poorly suited to answering that question. Of 401 datasets used to evaluate
22 time-series foundation models, only about 6\% had never appeared in any model's pre-training
corpus, and even a 0.1\% overlap can move the mean absolute percentage error by 8 to 29 percentage
points \citep{meyer2025leakage}. Foundation-model evaluations have also been criticised for comparing
against under-tuned classical baselines rather than the combinations that win forecasting competitions
\citep{makridakis2020m4}. This study therefore generates its own evaluation data. Under a known
data-generating process (DGP) the foundation model cannot have seen the series, because they did not
exist before the experiment, and the model families searched by AutoARIMA and AutoETS contain the true
process of the linear and exponential-smoothing DGPs. What the design excludes is contamination at the
level of the realisation; what it cannot exclude is familiarity at the level of the family, because
TimesFM-3 is pre-trained partly on synthetic series \citep{timesfm3blog}. Such familiarity cannot be
measured here, and its direction cannot be assumed: synthetic corpora contain autoregressive and seasonal
structure, resembling D1--D5, but also level shifts, piecewise trends and sparse spikes, resembling
D6--D8. No conclusion of this paper relies on its direction.

The aim of the study is exploratory: to describe, under controlled conditions, how the accuracy of
TimesFM-3 relative to classical methods varies across regimes, and to derive diagnostic baselines for
when it may be preferred and when it should not. It proposes no new estimator, test or theory. The
design can establish what TimesFM-3 does but not why: the 330 million parameters have no individual
interpretation and the pre-training corpus cannot be inspected, so explanations of its behaviour offered
here are interpretations consistent with the measurements, not findings.

The study makes four contributions. First, it evaluates a time-series foundation model on data that are
free of realisation-level contamination by construction, under a protocol frozen before any forecast was
produced, with every departure recorded. Second, it characterises relative accuracy by the worst case
across regimes as well as by the average, and shows that, among five foundation models, a bounded worst
case is specific to TimesFM-3 and confined to horizons of up to two years of monthly steps. Third, it
separates correct specification from estimability: the largest failures of automatic ARIMA and ETS arise
when two seasonal cycles do not suffice to fit a seasonal form. Fourth, it places the foundation models
against the published forecasts of the leading M4 submissions and against simple history-based benchmarks
for intermittent demand, and it records the licensing and cost constraints that bear on their use.
\sectionref{sec:timesfm} describes TimesFM-3, \sectionref{sec:design} the design and the pre-registered
hypotheses, \sectionref{sec:results} the pre-registered comparison, \sectionref{sec:posthoc} the post-hoc
analyses of the simulation, \sectionref{sec:realdata} the application to M4 and to data
observed after the training corpora, and \sectionref{sec:discussion}
the discussion.

\subsection{Related work}
\label{sec:related}

TimesFM \citep{das2024timesfm}, Chronos \citep{ansari2024chronos}, Moirai \citep{woo2024moirai},
TimeGPT \citep{garza2023timegpt} and Lag-Llama \citep{rasul2023lagllama} established that a single
pre-trained network can forecast unseen series zero-shot, and GIFT-Eval \citep{aksu2024gifteval} and
fev-bench \citep{shchur2025fevbench} supplied common benchmarks that include AutoARIMA, AutoETS and
seasonal naive among their baselines; newer models include MOMENT \citep{goswami2024moment}, TTM
\citep{ekambaram2024ttm}, Time-MoE \citep{shi2024timemoe}, Sundial \citep{liu2025sundial}, Toto
\citep{cohen2025toto} and Moirai 2.0 \citep{liu2025moirai2}. The comparison of statistical and machine-learning methods has a
longer history in this literature \citep{makridakis2018statistical}, and the critical literature on
foundation models has grown alongside: \citet{meyer2025leakage} on contamination, and
\citet{adler2026calibrated}, who found on six public datasets that the calibration error of five
foundation models grows with forecast length. The present study differs from the latter in using generated
rather than public data, in testing accuracy differences with multiplicity control rather than only
summarising calibration, and in evaluating TimesFM-3. Applied comparisons on real data are appearing in
several fields---\citet{chu2026influenza} compare six foundation models with SARIMA on influenza
surveillance data---but they cannot say why a model works, because the generating mechanism is unknown.

The design follows a pattern established in this journal, a controlled simulation over several DGPs
followed by an empirical application \citep{berger2025deep, chu2026intervals, lisi2025multiple}, and it
checks the realism of the simulated series in a feature space, as in \citet{kang2017instance} and
\citet{talagala2023metalearning}. AutoARIMA and AutoETS follow the Hyndman--Khandakar procedures
\citep{hyndman2008forecast, hyndman2002statespace}, Theta \citep{assimakopoulos2000theta} won M3
\citep{makridakis2000m3}, and the combination is included because combinations are the standard against
which competition entries are judged \citep{makridakis2000m3, makridakis2020m4}. Evaluation follows the
scale-free conventions of \citet{hyndman2006mase} and the M5 competition \citep{makridakis2022m5,
makridakis2022m5unc} and avoids the pitfalls catalogued by \citet{hewamalage2023evaluation}.

\section{The TimesFM-3 model}
\label{sec:timesfm}

\subsection{Architecture and pre-training data}
\label{sec:tfm-architecture}

TimesFM-3 is the third generation of the decoder-only TimesFM family \citep{das2024timesfm,
timesfm3blog}. Each input series is normalised by an iterative reversible instance normalisation, with
linear detrending of strongly trending inputs, and divided into non-overlapping patches of 32 time points,
each embedded as a token; 20 transformer layers with model dimension 1280 and 16 attention heads give
about 330 million parameters (the checkpoint loaded here has 330\,710\,976). Layers of causal attention
along time alternate with layers of attention across variates, so target series and covariates can be
processed jointly, and a contiguous patch masking scheme produces the whole horizon in one forward pass,
in output patches of 64 points. For each step the model returns nine quantiles, the 10th to the 90th
percentile, and its point forecast is the median \citep{timesfm3blog, timesfm3card, timesfm3repo}. These
details come from the developers' blog post and model card and from the configuration file of the
checkpoint (accessed 23 September 2026); none has yet been described in a peer-reviewed publication. The
pre-training corpus of more than one trillion time points combines the GIFT-Eval pre-training collection,
Wikipedia page views to November 2023, Google Trends queries to the end of 2022, and synthetic and
augmented series \citep{timesfm3card}.

\subsection{Checkpoint and configuration}
\label{sec:tfm-config}

The released weights are loaded with the developers' \proglang{Python} package (versions and revisions in
\sectionref{sec:repro}). The model is used zero-shot: it receives only the raw context, with no fine-tuning,
covariates or seasonal period, through \code{predict\_batch} with its default settings (no symmetric
averaging, no positivity constraint, 32-bit arithmetic). Its point forecast is the model's own median
output; the nine deciles are then sorted, which matters only where they cross (\sectionref{sec:checks}). A
context shorter than the model's working length is left-padded with zeros and masked, so at $n = 24$ the
model sees one partly filled patch; the package's maximum context of 15\,360 points truncates no series in
this study, and a series forecast alone or inside a batch of other lengths gives quantiles that agree to
within $2 \times 10^{-5}$. Because no quantile outside the 10th and 90th percentiles is available, the widest
interval that can be formed is the nominal 80\% interval.

\section{Design of the simulation study}
\label{sec:design}

\tableref{tab:ademp} summarises the design in the aims--data--estimands--methods--performance form
recommended for simulation studies by \citet{morris2019simulation}.

\begin{table}[!htbp]
\centering
\begin{threeparttable}
\caption{Structure of the simulation study.}
\label{tab:ademp}
\small
\begin{tabular}{>{\raggedright\arraybackslash}p{2.6cm}>{\raggedright\arraybackslash}p{11.4cm}}
\toprule
Element & Specification \\
\midrule
Aims & Describe how the accuracy of zero-shot TimesFM-3 relative to classical methods varies across
  known data-generating regimes; exploratory. \\[2pt]
Data-generating mechanisms & Main design: nine processes (\tableref{tab:dgps}) with fixed
  parameters, $n \in \{24, 48, 96, 200\}$, $H = 12$, 200 replications per scenario. Post hoc: parameters drawn
  per series from wide ranges in five versions (\sectionref{sec:departures}); the nine processes
  regenerated for $H = 48$ (\sectionref{sec:horizon48}). \\[2pt]
Estimands & For each method and scenario: the expected MASE over $h = 1, \dots, 12$ (and RMSSE, scaled
  pinball loss, 80\% coverage); for each pair of methods and scenario: the pseudo-median of the paired
  MASE difference. Summaries across scenarios (worst ratio, win counts) were defined post hoc. \\[2pt]
Methods & Seasonal naive, Theta, AutoETS, AutoARIMA, their combination and TimesFM-3 (pre-registered).
  Post hoc: DOTM, the M4 Comb benchmark, a second combination, a median combination, an Oracle ARIMA, the
  Croston family, the all-zero and history benchmarks on D8, TimesFM-2.5 (two settings), Chronos-Bolt,
  Chronos-2 and TiRex. \\[2pt]
Performance measures & Mean MASE with its Monte Carlo standard error; ratio to the best method in
  a scenario, with bootstrap intervals for summaries across scenarios; paired Wilcoxon tests with
  Hodges--Lehmann estimates and Benjamini--Hochberg control; coverage and width of 60\% and 80\%
  intervals. With 200 replications the Monte Carlo standard error of a paired MASE difference is about
  2\% of the mean (\sectionref{sec:effects}). \\
\bottomrule
\end{tabular}
\begin{tablenotes}[flushleft]\footnotesize
\item \textit{Note:} The mean MASE (descriptive) and the pseudo-median difference (inferential)
are different estimands and can disagree when the error distribution is skewed, as on the
intermittent process D8 (\sectionref{sec:effects}); both are reported.
\end{tablenotes}
\end{threeparttable}
\end{table}

\subsection{Data-generating processes}
\label{sec:dgps}

\begin{figure}[!htbp]
\centering
\includegraphics[width=\textwidth]{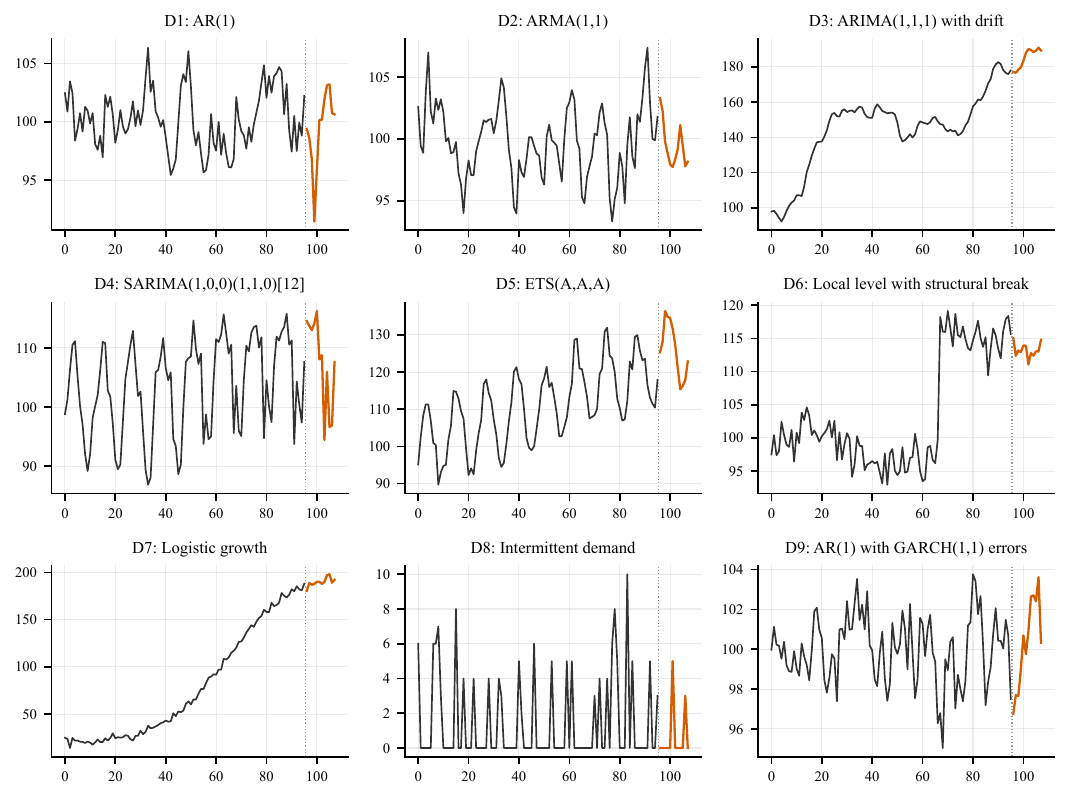}
\caption{One realisation of each data-generating process at $n = 96$. The context supplied to
every method is in grey and the held-out horizon in colour, separated by the dotted line.}
\label{fig:series}
\end{figure}

Nine processes were chosen to span the range from ``a classical model is exactly right'' to
``no classical model is right''. \tableref{tab:dgps} lists them. Each series is generated at
length $n + H$ with $H = 12$; the first $n$ observations form the context supplied to every
method, and the final 12 are held out as the truth. One realisation of each process is shown
in \figureref{fig:series}.

\begin{table}[!htbp]
\centering
\begin{threeparttable}
\caption{The nine data-generating processes and their parameters.}
\label{tab:dgps}
\small\setlength{\tabcolsep}{4pt}
\begin{tabular}{llll}
\toprule
ID & Data-generating process & Parameters & Correctly specified by \\
\midrule
D1 & AR(1)                        & $\phi = 0.7$                                & ARIMA \\
D2 & ARMA(1,1)                    & $\phi = 0.6$, $\theta = 0.4$                & ARIMA \\
D3 & ARIMA(1,1,1) with drift      & $\phi = 0.5$, $\theta = 0.3$, drift $0.3$   & ARIMA \\
D4 & SARIMA(1,0,0)(1,1,0)$_{12}$  & $\phi = 0.6$, $\Phi = 0.5$                  & ARIMA \\
D5 & ETS(A,A,A)                   & $\alpha = 0.3$, $\beta = 0.01$, $\gamma = 0.2$ & ETS \\
D6 & Local level with a break     & break at $0.7n$, jump $10\sigma$            & neither \\
D7 & Logistic growth              & $K = 180$, $r = 0.08$, floor $20$           & neither \\
D8 & Intermittent demand          & $p = 0.3$, size $1 + \mathrm{Pois}(4)$      & neither \\
D9 & AR(1) with GARCH(1,1) errors & $\phi = 0.6$, $a = 0.1$, $b = 0.85$         & ARIMA in mean \\
\bottomrule
\end{tabular}
\begin{tablenotes}[flushleft]\footnotesize
\item \textit{Note:} The last column states which classical family is correctly specified by
construction; for D6--D8 none is, and for D9 the conditional mean is correctly specified while
the conditional variance is not. Innovations are Gaussian with standard deviation $\sigma = 2$ ($\sigma = 3$
in D7; D8 has none). D1, D2 and D9 fluctuate about a mean of 100 after a burn-in of 300 steps (D9:
$\omega = 0.1$); D3 starts at 100 after its differences are burnt in; D4 starts from the cycle
$100 + 10\sin(2\pi k/12)$ plus noise, its seasonal differences burnt in; D5 starts from level 100, trend
0.2 and seasonal states $10\sin(2\pi k/12)$; in D6 the level is a random walk from 100 with innovation
standard deviation 0.5 and $\sigma$ is that of the observation noise; D7 has lower asymptote 20; in D8 a
demand occurs with probability $p$.
\end{tablenotes}
\end{threeparttable}
\end{table}

The parameter values in \tableref{tab:dgps} were fixed in the pre-registered protocol, each chosen so
that its process represents its regime unambiguously. The linear processes D1--D4 and the conditional
mean of D9 use coefficients no larger than 0.7 in absolute value, well inside the stationarity and
invertibility regions \citep{brockwell2016introduction}. The smoothing parameters of D5 lie inside the
usual region $0 < \beta < \alpha < 1$, $0 < \gamma < 1 - \alpha$ \citep{hyndman2008ets}, with a small trend
parameter so that the level stays positive. The break in D6 equals ten noise standard deviations, so the
comparison concerns adaptation to the break rather than its detection \citep{pesaran2007breaks}; D7
follows the logistic curve, the standard model of saturating growth \citep{meade2006diffusion}. In D8 the
mean interval between demands is $1/0.3 \approx 3.3$ periods and the squared coefficient of variation of
the sizes is $4/25 = 0.16$, so against the cut-offs of \citet{syntetos2005categorization} the process is
intermittent rather than lumpy, the regime for which Croston's method was developed \citep{croston1972}.
The GARCH(1,1) errors of D9 \citep{bollerslev1986garch} have persistence $a + b = 0.95$, strongly
persistent but covariance-stationary \citep{engle1986persistence}, a standard benchmark specification in
volatility forecasting \citep{hansen2005garch}. Every series has a level of about 100, or a positive lower
asymptote in D7, so percentage errors are defined except where zeros are intrinsic (D8); the trend
parameter of D5 and the asymptote of D7 were adjusted for this reason before any forecast was produced.

Three design choices need a word. The break in D6 is placed inside the context, at $0.7n$: a break in the
held-out window is unforecastable by any method, whereas one in the context asks whether a method
forecasts from the new level or reverts to the old one. The inflection of D7 sits at $0.6(n+H)$, so the
held-out window lies in the saturating arm, where a method that fits a linear drift overshoots. And the
shortest length, $n = 24$, leaves a seasonal model only two cycles from which to estimate its seasonal
structure, which separates whether a model is correctly \emph{specified} from whether it is
\emph{estimable}. Two hundred replications per (DGP, length) scenario give $9 \times 4 \times 200 = 7\,200$
series, each seeded from its own identifiers.

\subsection{Forecasting methods}
\label{sec:methods}

No method is tuned by hand; each classical model uses its own automatic selection, the comparison a
practitioner faces. The classical models are fitted with \pkg{StatsForecast} \citep{statsforecast} at its
default settings, stated because the results depend on them. AutoARIMA searches stepwise over $p, q \le 5$
and $P, Q \le 2$ with $p + q + P + Q \le 5$, chooses $d \le 2$ and $D \le 1$ by unit-root tests, allows a
mean or drift and selects by AICc; AutoETS searches the full error--trend--season space (\code{ZZZ}) with
damped and undamped trends by AICc; Theta is the standard theta method with classical multiplicative
seasonal adjustment \citep{hyndman2021fpp}. The equal-weight combination of Theta, AutoETS and AutoARIMA
is included because the classical state of the art is a combination \citep{makridakis2020m4,
petropoulos2022theory}. Its point forecast is the mean of the members' forecasts and its quantiles the
decile-by-decile mean of theirs (vincentization), which gives a narrower distribution than a mixture when
the members disagree. Every classical method receives the true seasonal period ($m = 12$ on D4 and D5,
$m = 1$ elsewhere), whereas TimesFM-3 receives only the raw context; this reflects ordinary practice---an
analyst usually knows that data are monthly---and \sectionref{sec:departures} examines the alternative of
testing for seasonality.

\subsection{Evaluation measures and statistical inference}
\label{sec:evaluation}

Point accuracy is measured mainly by the mean absolute scaled error \citep[MASE;][]{hyndman2006mase}.
With context $y_1, \dots, y_T$ ($T = n$), horizon $H$ and point forecasts $\hat{y}_{T+h}$,
\begin{equation}
\label{eq:mase}
\text{MASE} = \frac{\frac{1}{H}\sum_{h=1}^H |y_{T+h} - \hat{y}_{T+h}|}{\frac{1}{T-m}\sum_{t=m+1}^T |y_t - y_{t-m}|},
\end{equation}
so the scale is the in-sample seasonal-naive error for the seasonal processes and the naive error for the
others ($m = 12$ throughout for the M4 tier, as in M4). The root mean squared scaled error (RMSSE) replaces
absolute by squared errors in both numerator and denominator and takes the square root, and sMAPE is the
symmetric percentage error of the M4 competition. MAPE is undefined when an actual value is zero, which
happens only on D8, and is reported only in the Supporting Information (Section~S1)
\citep{hyndman2006mase, kolassa2016count}. With $m = 12$ and $n = 24$ the MASE denominator rests on 12
seasonal differences: its coefficient of variation across replications is 0.33 on D4 (0.09 at $n = 200$)
and 0.34 to 0.44 on D5 at every length, so absolute MASE levels in those scenarios are imprecise, whereas
the paired ratios used for inference are much less affected. Probabilistic accuracy is measured by a strictly proper scoring
rule \citep{gneiting2007scoring}, the pinball loss over the nine deciles $\tau \in \{0.1, \dots, 0.9\}$
scaled by the same denominator,
\begin{equation}
\label{eq:spl}
\text{SPL} = \frac{\frac{1}{9H} \sum_{\tau} \sum_{h=1}^H \max\{\tau(y_{T+h} - \hat{q}^{(\tau)}_{T+h}),\, (1-\tau)(\hat{q}^{(\tau)}_{T+h} - y_{T+h})\}}{\frac{1}{T-m}\sum_{t=m+1}^T |y_t - y_{t-m}|},
\end{equation}
where $\hat{q}^{(\tau)}_{T+h}$ is the $\tau$-quantile forecast, together with the coverage and width of
the nominal 60\% and 80\% intervals \citep{gneiting2007calibration}. Wider intervals cannot be formed from
TimesFM-3's nine deciles, so $[q_{0.1}, q_{0.9}]$ is the widest interval compared; classical quantiles are
taken from intervals at levels 20, 40, 60 and 80, which map onto the same deciles.

Within each (DGP, length) scenario and horizon slice ($h = 1$, $1$--$6$ and $1$--$12$), TimesFM-3 is compared
with each classical method by a paired Wilcoxon signed-rank test \citep{wilcoxon1945} on per-replication
MASE differences, with the Hodges--Lehmann estimate \citep{hodges1963} and the median MASE ratio as effect
sizes, and the Benjamini--Hochberg procedure at a false discovery rate of 0.05 within each opponent family
of 108 tests (36 scenarios and three slices), as pre-registered \citep{benjamini1995fdr}; the results report the
$h = 1$--12 slice. The Diebold--Mariano test \citep{diebold1995comparing} is not used: the replications are
independent, so there is no serial dependence for it to model, and in the real-data tier three rolling
origins per series leave it without power even with small-sample corrections \citep{harvey1997testing}. The
pre-registered protocol had reserved it for that tier; the departure is recorded.

\subsection{Protocol and pre-registered hypotheses}
\label{sec:protocol}

The protocol was written and frozen on 8 September 2026, before any forecast was produced; it fixed the
processes, lengths, methods, measures and tests above and five hypotheses. \textbf{H1}: on D1--D5, where a
classical family is correctly specified, AutoARIMA or AutoETS is more accurate than TimesFM-3, with the gap
largest at $n = 24$. \textbf{H2}: on D6--D8, where none is, TimesFM-3 is more accurate than every classical
method. \textbf{H3}: on D9 the two families are indistinguishable in the mean, but the classical intervals
under-cover. \textbf{H4}: the combination beats every individual classical method and is the hardest
opponent for TimesFM-3. \textbf{H5}: the 80\% intervals of both families under-cover, TimesFM-3's more at
longer horizons. The protocol stated no decision rules, so the verdicts in \sectionref{sec:hypotheses} are
judgements on the pattern of the pre-registered tests. Everything in \sectionref{sec:posthoc} and \sectionref{sec:realdata}, except the rolling-origin evaluation
of \sectionref{sec:m4results}, was added afterwards and is logged as a departure.

\section{Pre-registered comparison}
\label{sec:results}

\begin{figure}[!htbp]
\centering
\includegraphics[width=\textwidth]{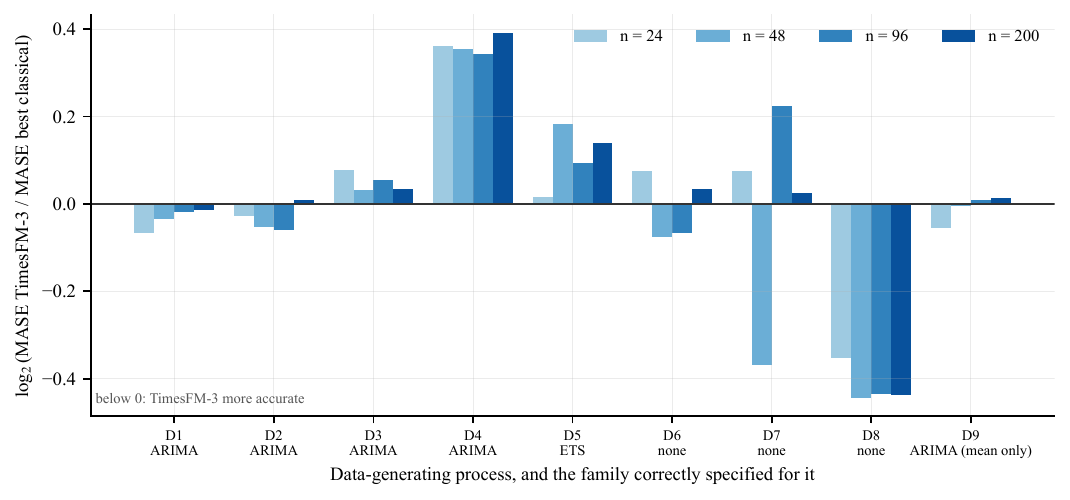}
\caption{Accuracy of TimesFM-3 relative to the best classical method in each scenario, on a
$\log_2$ scale. Bars below zero mark scenarios where TimesFM-3 is more accurate. The label under
each process names the family correctly specified for it by construction.}
\label{fig:ratio}
\end{figure}

This section reports the forecasts and tests fixed by the protocol. The summaries across scenarios used
below---the worst ratio to the best method of a scenario and the count of scenarios won---were defined after the
protocol and are descriptive summaries of the pre-registered forecasts. \figureref{fig:ratio} shows
TimesFM-3 relative to the best classical method in each of the 36 (process, length) scenarios; the mean MASE of
every method in every scenario, with Monte Carlo standard errors, is in the Supporting Information (Table~S2).

\subsection{Overall accuracy and pairwise comparisons}
\label{sec:overall}

\begin{figure}[!htbp]
\centering
\includegraphics[width=\textwidth]{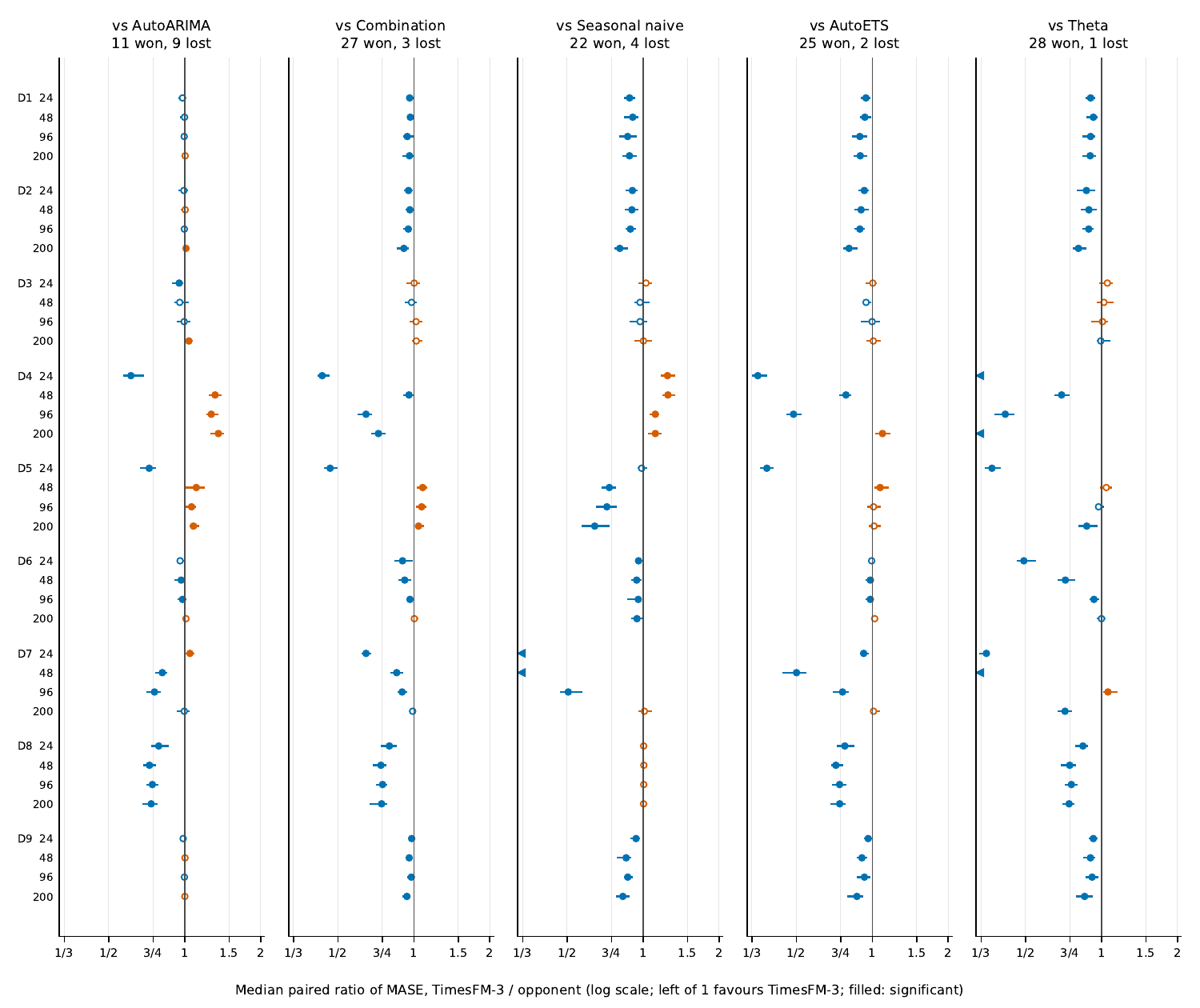}
\caption{Pairwise comparisons of TimesFM-3 with each classical method, scenario by scenario: median of the paired
per-series ratios of MASE over $h = 1, \dots, 12$, with 95\% bootstrap intervals, on a logarithmic scale.
Filled markers are significant after Benjamini--Hochberg correction within each opponent family; points
left of 1 favour TimesFM-3. Panel titles give the significant comparisons won and lost by TimesFM-3.
Arrowheads mark medians beyond the plotted range.}
\label{fig:forest}
\end{figure}

TimesFM-3 has the lowest mean MASE in 16 of the 36 scenarios and the classical methods in 20: AutoARIMA 10,
seasonal naive 4, and two each for Theta, AutoETS and the combination. Against the best classical method of
each scenario, 18 differences survive Benjamini--Hochberg correction, 7 favouring TimesFM-3 and 11 the classical
method; because that comparator is selected on the same replications, these $p$-values are descriptive
only. The inferential claims rest on the 180 comparisons with a fixed opponent (\figureref{fig:forest}):
132 are significant, 113 favouring TimesFM-3; pooling the tests into a single family changes these counts
by at most two (Supporting Information, Section~S6). \textbf{AutoARIMA is the only classical method that
TimesFM-3 does not clearly beat} (11 comparisons won, 9 lost): 16 of the 36 intervals against it straddle 1, and AutoARIMA's wins are mainly
on the seasonal processes from $n = 48$, whereas against the other methods,
the combination included, most intervals lie wholly on the side of TimesFM-3.

\subsection{Effect sizes and sensitivity to the choice of test}
\label{sec:effects}

The median Hodges--Lehmann shifts in MASE across the 36 scenarios are $-0.010$ against AutoARIMA, $-0.091$
against the combination, $-0.133$ against seasonal naive, $-0.183$ against AutoETS and $-0.227$ against
Theta (negative favours TimesFM-3). The Monte Carlo standard error of the paired difference between
TimesFM-3 and AutoARIMA is 0.8\% to 4.8\% of AutoARIMA's mean MASE (median 2.1\%), so differences smaller than
about 6\% are unlikely to be detected in a scenario: the split against AutoARIMA sits at the resolution of the
design and does not show that the two methods are equivalent. A paired $t$-test agrees with the Wilcoxon
test in 163 of 180 comparisons (90.6\%). The largest block of disagreements is on the intermittent process D8
against seasonal naive: the $t$-test finds TimesFM-3 better at every length ($p < 0.0001$), the signed-rank
test finds no difference ($p = 0.11$ to $0.92$). The mean difference is driven by a minority of series on
which seasonal naive fails badly, whereas on 64.5--72.0\% of series TimesFM-3 is marginally less
accurate---the two tests estimate different quantities of a heavily skewed error distribution.

\subsection{Correct specification versus estimability}
\label{sec:estimability}

On D1 and D2 TimesFM-3 is at least as accurate as the correctly specified AutoARIMA at every length except
D2 at $n = 200$, the only significant difference: correct specification bought AutoARIMA nothing
detectable on short and moderate series. The seasonal processes show why most clearly. With two complete
cycles ($n = 24$), AutoETS scores a mean MASE of \textbf{3.03 on D5, data an ETS process generated}, against
1.14 for seasonal naive and 1.15 for TimesFM-3: the correctly specified model has 2.7 times the error of the
naive rule. The failure is not an over-fitted seasonal model but the absence of one: given $m = 12$ and 24
observations, AutoETS chose a seasonal form for 8.5\% of the D4 series and none of the D5 series, and
AutoARIMA for none (Supporting Information, Table~S11), whereas seasonal naive applies the known period
without estimating anything. The operative distinction is therefore whether a model is \emph{estimable}
from the series at hand, not only whether it is correctly specified.

\tableref{tab:robust} quantifies the resulting asymmetry with each method's mean MASE as a multiple of the
best in the same scenario---a relative measure, because D3 is hard for every method and would dominate absolute
worst cases. \textbf{TimesFM-3 is never worse than $1.31\times$ the best method of a scenario}, and within 0.8\% of
it in the median scenario, whereas every classical method is at least $2.2\times$ worse somewhere: AutoARIMA
$2.22\times$ and AutoETS $3.50\times$ (both on D4 at $n = 24$), Theta $4.69\times$, seasonal naive $5.65\times$.
Leaving out the two scenarios with two seasonal cycles, TimesFM-3's worst ratio is unchanged and AutoARIMA's
falls to 1.62 (on D7 at $n = 96$). The worst ratio depends on the methods compared, notably on seasonal naive
with the known period (\sectionref{sec:departures}).

\begin{table}[!htbp]
\centering
\begin{threeparttable}
\caption{Mean MASE relative to the best method in the same scenario, summarised over the 36 scenarios.}
\label{tab:robust}
\small\setlength{\tabcolsep}{4pt}
\begin{tabular}{lrrrlrl}
\toprule
 & & & \multicolumn{2}{c}{All 36 scenarios} & \multicolumn{2}{c}{Without the two-cycle scenarios} \\
\cmidrule(lr){4-5}\cmidrule(lr){6-7}
Method & Median & Mean & Worst & Scenario & Worst & Scenario \\
\midrule
Seasonal naive & 1.200 & 1.470 & 5.65 & D7 $n=48$ & 5.65 & D7 $n=48$ \\
Theta & 1.251 & 1.614 & 4.69 & D4 $n=200$ & 4.69 & D4 $n=200$ \\
AutoETS & 1.194 & 1.366 & 3.50 & D4 $n=24$ & 2.72 & D4 $n=96$ \\
AutoARIMA & 1.032 & 1.137 & 2.22 & D4 $n=24$ & 1.62 & D7 $n=96$ \\
Combination & 1.108 & 1.273 & 3.00 & D4 $n=24$ & 2.08 & D4 $n=96$ \\
\textbf{TimesFM-3} & \textbf{1.008} & \textbf{1.054} & \textbf{1.31} & D4 $n=200$ & \textbf{1.31} & D4 $n=200$ \\
\bottomrule
\end{tabular}
\begin{tablenotes}[flushleft]\footnotesize
\item \textit{Note:} Each entry divides a method's mean MASE by the lowest mean MASE in the same (process,
length) scenario, so a value of 1.00 means that the method was the best in that scenario; the mean is arithmetic.
The two-cycle scenarios are D4 and D5 at $n = 24$. Geometric means and 90th percentiles are in the Supporting
Information, Table~S9.
\end{tablenotes}
\end{threeparttable}
\end{table}

\subsection{Seasonal and intermittent regimes}
\label{sec:regimes}

\textbf{Strong seasonality with sufficient data (D4).} At $n = 48$, 96 and 200 AutoARIMA's mean MASE is
21.8\%, 21.1\% and 23.7\% below TimesFM-3's (adjusted $p < 0.0001$), and D5 shows the same more mildly (4\%
to 12\%). Once at least four cycles are observed, AutoARIMA recovers the seasonal structure and TimesFM-3
does not match it; seasonal naive, which estimates nothing, is also more accurate than TimesFM-3 on D4 at
every length (1.005 to 1.053 against 1.124 to 1.321). With two cycles the ranking among the others reverses:
TimesFM-3 beats AutoARIMA by 42\% (1.321 against 2.284). Theta deteriorates on D4 as the series lengthens
(1.94 at $n = 48$, 4.16 at $n = 200$): D4 is seasonally integrated, so its seasonal pattern drifts, whereas
classical decomposition estimates one fixed pattern from the whole history.

\textbf{Intermittent demand (D8).} On MASE TimesFM-3 reduces error against the best general-purpose
method by 21.7\% to 26.5\% at the four lengths (adjusted $p < 0.0001$) and has the lowest pinball loss of
the pre-registered methods, but it \emph{loses} on RMSSE at every length (Supporting Information,
Table~S6); sMAPE, which scores a zero forecast of a zero outcome as perfect, is not informative here. About
70\% of D8's observations are zero: MASE is minimised by the conditional median, which is zero or near it,
RMSSE by the conditional mean, which is positive, so a median-like forecast wins MASE and loses RMSSE
\citep{kolassa2016count, gneiting2011making}. \sectionref{sec:d8-extended} shows that both advantages are
matched by simple benchmarks built from the history.

Across horizons ($h = 1$, $1$--$6$ and $1$--$12$) every method degrades and no regime-average ranking
reverses (Supporting Information, Figure~S1).

\subsection{Coverage of intervals and the combination}
\label{sec:coverage}

\begin{figure}[!htbp]
\centering
\includegraphics[width=\textwidth]{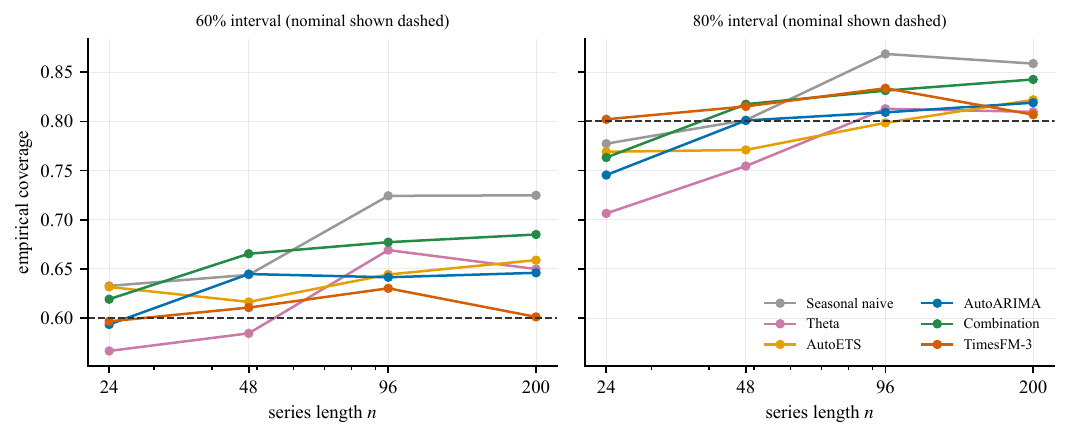}
\caption{Empirical coverage of the nominal 60\% and 80\% intervals, averaged over all nine
processes, against series length. The nominal level is dashed. TimesFM-3's 80\% coverage stays
within 0.802--0.834 across every length; the classical methods under-cover at $n = 24$ and reach or exceed nominal from $n = 96$.}
\label{fig:coverage}
\end{figure}

Averaged over the processes (\figureref{fig:coverage}), TimesFM-3's 80\% coverage stays between 0.802 and
0.834 at every length, with the smallest mean absolute deviation from nominal across lengths of any method, and within the
12-step horizon it does not fall with the horizon (0.786 at $h = 1$, 0.814 over $h = 1$--12). Averaging lets
over- and under-coverage cancel, however: the mean per-scenario absolute deviation from nominal is 0.047 for
TimesFM-3, 0.066 for AutoARIMA, 0.089 for the combination, 0.101 for AutoETS, 0.127 for Theta and 0.144 for
seasonal naive, and TimesFM-3's per-scenario 80\% coverage ranges from 0.583 to 0.921. No method delivers
reliable 80\% intervals on all nine processes. The classical intervals rest on Gaussian innovations,
violated by construction on D8 and D9. An interval score, which rewards narrow intervals and penalises
outcomes outside them (80\% interval, scaled as MSIS), puts TimesFM-3 first in 22 of the 36 scenarios
(Supporting Information, Table~S12). On the structural break D6 every classical method covers 0.966 to
0.999, because the break inflates the estimated variance and the intervals become very wide (scaled widths
4.8 to 10.4); TimesFM-3 covers 0.828 to 0.889 with intervals about half as wide (3.0 to 5.1) and has the
lowest interval score at every length. Against the pre-registered mean combination TimesFM-3 wins 27 of 30
significant comparisons.

\subsection{Assessment of the pre-registered hypotheses}
\label{sec:hypotheses}

\tableref{tab:hypotheses} sets the pre-registered expectations against the evidence. \textbf{None of the five
survives unqualified}: H1 and H5 are not supported, H4 is contradicted, and H2 and H3 are partially
supported.

\begin{table}[!htbp]
\centering
\begin{threeparttable}
\caption{Pre-registered hypotheses and the evidence of the pre-registered comparison.}
\label{tab:hypotheses}
\small
\begin{tabular}{>{\raggedright\arraybackslash}p{0.6cm}>{\raggedright\arraybackslash}p{4.4cm}>{\raggedright\arraybackslash}p{2.0cm}>{\raggedright\arraybackslash}p{6.6cm}}
\toprule
 & Expectation & Verdict & Evidence \\
\midrule
H1 & Correct classical models win on D1--D5, most at $n = 24$ & Not supported & At $n = 24$ the gap runs the
  other way; the classical side wins mainly where the model is also estimable (D4, D5 at $n \geq 48$) \\[2pt]
H2 & TimesFM-3 wins on D6--D8 & Partially & Mildly on breaks (D6); depends on length on D7; on D8 only against
  the classical methods (\sectionref{sec:d8-extended}) \\[2pt]
H3 & D9: tie in the mean, classical intervals under-cover & Partially & Tie in the mean; only AutoARIMA
  under-covers \\[2pt]
H4 & The combination beats every classical method and is the hardest opponent & Contradicted & AutoARIMA has the
  lower mean MASE and is the harder opponent (11--9 against 27--3) \\[2pt]
H5 & Both families under-cover, TimesFM-3 more at longer horizons & Not supported & TimesFM-3 is close to nominal
  on average and does not worsen within 12 steps; the classical methods reach or slightly exceed nominal from $n = 96$ \\
\bottomrule
\end{tabular}
\begin{tablenotes}[flushleft]\footnotesize
\item \textit{Note:} The protocol stated no decision rules; verdicts are judgements on the pattern of the
pre-registered tests at $h = 1, \dots, 12$.
\end{tablenotes}
\end{threeparttable}
\end{table}

\section{Post-hoc analyses of the simulation}
\label{sec:posthoc}

Everything in this section was designed after the results of \sectionref{sec:results} were known and is
logged as a departure from the protocol.

\subsection{Precision of the summaries}
\label{sec:mcse}

The Monte Carlo standard error of a scenario's mean MASE is 1.5\% to 6.2\% of the mean (median 3.2\%;
Supporting Information, Table~S2). Bootstrapping the replications within each scenario (2\,000 draws, paired
across methods), the worst ratio of TimesFM-3 to the scenario-best method is 1.31 (95\% interval 1.28 to 1.36)
and that of AutoARIMA, the smallest among the classical methods, 2.22 (2.07 to 2.38); their difference has
an interval of 0.75 to 1.05. Because the best method of a scenario is chosen on the same replications, the
statistic could be biased upwards; choosing it on one half of the replications and estimating the ratio on
the other gives 1.32, so the bias is small. These intervals reflect Monte Carlo error only, not the choice
of processes and parameters, which \sectionref{sec:departures} varies. The count of scenarios won is less
stable---TimesFM-3's 16 has an interval of 13 to 20---and no conclusion rests on it. Simultaneous
bootstrap bands over the 36 scenarios (max-$t$), which control the error of the whole set of intervals, are
about 1.9 times wider than the per-scenario intervals of \figureref{fig:forest}; against AutoARIMA they
separate 8 scenarios in favour of TimesFM-3 and 4 against it, and against every other opponent they keep
the direction of the pattern (Supporting Information, Table~S17).

\subsection{Implementation and configuration checks}
\label{sec:checks}

\textbf{Order selection.} An \emph{Oracle ARIMA}, which knows the true orders and estimates only the
parameters by Gaussian maximum likelihood, beats TimesFM-3 in 15 of the 16 ARIMA scenarios (for example 1.351
against 1.417 on D1 at $n = 24$), and on D1--D3 at $n = 24$ AutoARIMA pays a 9.9\% to 13.9\% penalty for
selecting the orders; the short-sample deficit of the automatic method is thus largely one of model
selection. On D4 at $n = 24$ most oracle fits did not converge (156 of 200, used as returned), and that scenario
is not interpreted.

\textbf{Implementation.} The D4 and D5 scenarios were refitted with \code{auto.arima}, \code{ets} and
\code{thetaf} of the \proglang{R} package \pkg{forecast} \citep{hyndman2008forecast}, the reference
implementation of the Hyndman--Khandakar procedures (Supporting Information, Table~S13). R reproduces every
failure discussed above---AutoARIMA 2.87 and AutoETS 3.95 on D4 at $n = 24$, AutoETS 3.30 on D5 at $n = 24$,
Theta 4.09 on D4 at $n = 200$---and, like \pkg{StatsForecast}, almost never fits a seasonal model with two
cycles. With additive instead of multiplicative decomposition Theta fails the same way on D4 (3.83 against
3.95 at $n = 200$ in 60 replications). Where both implementations choose seasonal forms they agree closely,
with one exception: on D4 at $n = 96$ \pkg{StatsForecast}'s AutoETS has a mean MASE of 2.41 against 1.04 for R,
because its optimiser settles on a near-zero seasonal smoothing parameter where R finds a high one. That scenario
is an implementation failure; with R's value there, AutoETS's worst ratio outside the two-cycle scenarios would
be 2.13, on D7 at $n = 48$.

\textbf{Combination rule.} Because the mean combination is not robust to a failing member such as Theta, a
\emph{median} combination of the same three methods was tried, in the spirit of \citet{petropoulos2020scum};
it is slightly worse (mean MASE 1.712 against 1.656, TimesFM-3 1.439), and TimesFM-3 still wins 25 of 29
significant comparisons. Under absolute error the mean combination can never be worse than the average of its
members, by convexity, a guarantee the median does not carry.

\textbf{TimesFM-3's configuration and output.} The developers' benchmark evaluator averages the forecasts of
a series and its negation and constrains non-negative series to non-negative forecasts. With these settings
TimesFM-3's worst ratio falls from 1.31 to 1.27; no conclusion changes. The output is a $12 \times 9$ array of
deciles whose raw median equals the point forecast exactly; quantile crossing occurs in 32\% to 46\% of steps
on D8 and essentially never elsewhere. Given the month of the year as past-and-future covariates
($\sin 2\pi t/12$, $\cos 2\pi t/12$) on D4 and D5---the information the classical methods receive through
their period---TimesFM-3's mean MASE rises in seven of the eight scenarios (for example 1.124 to 1.194 on D4 at
$n = 96$), so supplying the period in this form did not close AutoARIMA's seasonal advantage. Using the mean
of the deciles instead of the median as point forecast changes no process by more than 0.01 except D6 (0.835
to 0.875) and D8.

\subsection{Stronger classical benchmarks and other foundation models}
\label{sec:benchmarks}

\begin{figure}[!htbp]
\centering
\includegraphics[width=0.85\textwidth]{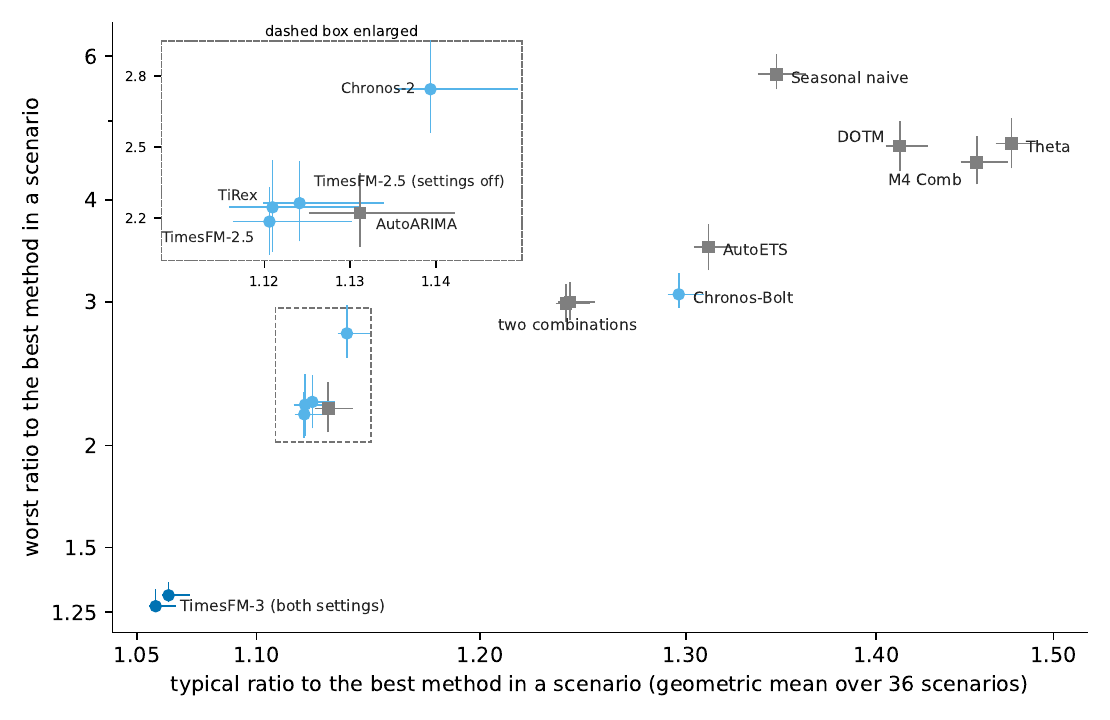}
\caption{Typical against worst-case accuracy of the fifteen methods of the main design: geometric mean over
the 36 scenarios of each method's ratio to the best method in a scenario, and the largest such ratio, with 95\%
bootstrap intervals. Circles are foundation models, squares classical methods; lower and further left is
better on both counts.}
\label{fig:scatter}
\end{figure}

Nine configurations were added to the six pre-registered methods (\figureref{fig:scatter}; Supporting
Information, Table~S10): dynamic optimised Theta \citep[DOTM;][]{fiorucci2016dotm}; the Comb benchmark of
M4, the mean of simple, Holt and damped exponential smoothing on seasonally adjusted data
\citep{makridakis2020m4}; a combination of AutoETS, AutoARIMA and DOTM; TimesFM-3 with its evaluator
settings; TimesFM-2.5 \citep{timesfm25card}, run with its evaluator-style settings and, to match the main
TimesFM-3 runs, with flip invariance and positivity switched off; Chronos-Bolt (Base, 205 million
parameters) \citep{ansari2024chronos, chronosboltcard}; and Chronos-2 \citep{ansari2025chronos2} and TiRex
\citep{auer2025tirex}, with 120 and 35 million parameters. All foundation models were run zero-shot with the
raw context only.

The stronger classical benchmarks do not change the picture. DOTM and the Comb benchmark share Theta's
failure on D4, where a decomposition with fixed seasonal indices meets a seasonal pattern that evolves, and
replacing Theta by DOTM leaves the combination's worst ratio at 2.99. TimesFM-3 wins 28 of 29 significant
comparisons against DOTM and all 29 against the Comb benchmark. \textbf{Among the five foundation models, the
robustness belongs to TimesFM-3 alone}: TimesFM-2.5 has a worst ratio of 2.18 (on D7 at $n = 24$), and 2.26
with its settings switched off, so the difference is not one of configuration; TiRex has 2.24, Chronos-2 2.75
and Chronos-Bolt 3.06, against 2.22 for AutoARIMA. In the median scenario every foundation model but Chronos-Bolt
is within 2\% to 4\% of the best, and TimesFM-2.5 is the best method in 7 scenarios, more than any model but
AutoARIMA (8). The newer models do not clearly beat AutoARIMA either (Chronos-2 10--10, TiRex 9--12).
Friedman tests with Nemenyi critical differences \citep{demsar2006statistical} within each regime agree:
TimesFM-3 is tied for the best mean rank on the linear processes (with TimesFM-2.5, TiRex and AutoARIMA)
and on the break and saturation processes, and AutoARIMA has the best mean rank on the seasonal processes.
Averaged over the design, the 80\% intervals of TimesFM-3 cover 0.814 (Monte Carlo SE 0.003), TimesFM-2.5
0.812, Chronos-2 0.789, TiRex 0.842, Chronos-Bolt 0.834 and AutoARIMA 0.794.

\subsection{Intermittent demand re-examined}
\label{sec:d8-extended}

Six purpose-built intermittent-demand methods were added: Croston's classic and optimised variants, the
Syntetos--Boylan approximation \citep{syntetos2005accuracy}, ADIDA \citep{nikolopoulos2011adida}, IMAPA
\citep{petropoulos2015imapa} and TSB \citep{teunter2011tsb}, the last with \pkg{StatsForecast}'s fixed
constants $\alpha_d = \alpha_p = 0.2$ (Supporting Information, Table~S7). TimesFM-3 beats all six on MASE at
every length (24/24 significant; median ratios 0.70 to 0.82), while the Croston family has the lower RMSSE
(0.694--0.769, against 0.774--0.806 for TimesFM-3). Because MASE rewards the conditional median, which on D8
is near zero, the trivial all-zero forecast also belongs in the comparison, and because D8 is independent
over time, so do benchmarks that use the empirical distribution of the history
\citep{willemain2004bootstrap}: its mean as point forecast and its deciles as predictive distribution.
\tableref{tab:d8ext} adds them, with the scaled mean error (bias) and periods in stock
\citep[PIS;][]{wallstrom2010pis}, which accumulates forecast errors as a stock position would.

\begin{table}[!htbp]
\centering
\begin{threeparttable}
\caption{Intermittent demand (D8) re-examined with simple benchmarks, bias and stock measures.}
\label{tab:d8ext}
\small\setlength{\tabcolsep}{4pt}
\begin{tabular}{lccccc}
\toprule
Method & MASE & RMSSE & sME & sPIS & SPL \\
\midrule
All-zero forecast & 0.668--0.718 & 0.785--0.825 & $-0.72$ to $-0.67$ & $-91$ to $-81$ & -- \\
History: mean; empirical deciles & 0.924--0.975 & 0.692--0.744 & $-0.07$ to $-0.02$ & $-13$ to $-3$ & 0.282--0.314 \\
TimesFM-3 & 0.683--0.780 & 0.774--0.806 & $-0.67$ to $-0.57$ & $-81$ to $-74$ & 0.288--0.325 \\
TimesFM-3, mean of deciles & 0.915--1.002 & 0.694--0.760 & $-0.06$ to $-0.01$ & $-10$ to $-4$ & 0.288--0.325 \\
TimesFM-2.5 & 0.678--0.770 & 0.777--0.806 & $-0.67$ to $-0.59$ & $-81$ to $-74$ & 0.289--0.324 \\
Chronos-Bolt & 0.708--0.786 & 0.758--0.789 & $-0.59$ to $-0.54$ & $-72$ to $-65$ & 0.294--0.328 \\
Chronos-2 & 0.681--0.750 & 0.776--0.808 & $-0.66$ to $-0.63$ & $-81$ to $-76$ & 0.290--0.323 \\
TiRex & 0.683--0.776 & 0.771--0.800 & $-0.67$ to $-0.58$ & $-81$ to $-73$ & 0.289--0.324 \\
Croston--SBA & 0.921--1.025 & 0.694--0.763 & $-0.04$ to $0.05$ & $-7$ to $2$ & -- \\
ADIDA & 0.926--0.972 & 0.700--0.753 & $-0.09$ to $-0.02$ & $-16$ to $-5$ & -- \\
TSB & 0.916--0.971 & 0.716--0.764 & $-0.10$ to $0.00$ & $-17$ to $-1$ & -- \\
AutoARIMA & 0.924--0.996 & 0.698--0.765 & $-0.05$ to $-0.03$ & $-12$ to $-3$ & 0.341--0.380 \\
\bottomrule
\end{tabular}
\begin{tablenotes}[flushleft]\footnotesize
\item \textit{Note:} Process D8, 200 replications per length; ranges over the four lengths $n \in \{24, 48, 96, 200\}$. sME: mean error (forecast minus actual) scaled by the MASE denominator; negative values mean under-forecasting. sPIS: periods in stock \citep{wallstrom2010pis}, cumulated over the horizon and scaled by the mean demand of the context; large negative values mean persistent stock-outs. SPL: scaled pinball loss over the nine deciles, defined only for methods with a predictive distribution. TimesFM-3, mean of deciles, uses the average of its nine deciles as point forecast. History: the mean of the context as point forecast and the empirical deciles of the context as predictive distribution; the median of the context is zero or near it, so its point forecast would repeat the all-zero row.
\end{tablenotes}
\end{threeparttable}
\end{table}

\textbf{The all-zero forecast has the lowest MASE at every length} (0.668 to 0.718), below TimesFM-3's
median forecast (0.683 to 0.780). The MASE advantage of \sectionref{sec:regimes} is thus the advantage of a
forecast close to zero: its bias ($-0.67$ to $-0.57$ MASE units) and PIS ($-81$ to $-74$) are close to those
of the all-zero forecast and imply persistent stock-outs. With the mean of its deciles as point forecast,
TimesFM-3 has an RMSSE of 0.694 to 0.760 and a bias within 0.06 of zero, level with the best Croston-family
method and with the mean of the history (0.692 to 0.744). Its pinball loss, 0.288 to 0.325, is about 15\%
below that of the classical methods (0.341 to 0.380), but their quantiles come from Gaussian intervals that
put probability on negative demand; the empirical deciles of the history score 0.282 to 0.314, slightly
lower than TimesFM-3 at every length. \textbf{On intermittent demand, therefore, TimesFM-3 shows no advantage
over simple benchmarks}: it behaves much like the empirical distribution of the history, and so do the other
foundation models (\tableref{tab:d8ext}). A model built for intermittent demand, iETS with automatic
occurrence selection \citep{svetunkov2023iets}, has a pinball loss of 0.283 to 0.306, lower than
TimesFM-3's at every length (paired Wilcoxon $p < 0.001$) and level with the empirical deciles: it selects
a fixed occurrence probability for 791 of 797 series, which brings its predictive distribution close to
that of the history. Negative-binomial and TSB-type predictive distributions fitted to the context do
worse than the empirical deciles (Supporting Information, Table~S15). The result concerns a process that
is independent over time, the
case most favourable to history-based benchmarks.

\subsection{Randomised parameters and departures from the assumptions}
\label{sec:departures}

Every series now draws its own parameters uniformly from wide ranges containing the fixed values of
\tableref{tab:dgps} (Supporting Information, Table~S8): autoregressive coefficients up to 0.95,
intermittency from lumpy ($p = 0.1$) to nearly continuous ($p = 0.6$), breaks of three to fifteen noise
standard deviations in either direction. Four versions use the same 7\,200 parameter draws (200 per process
and length): \emph{random parameters} with Gaussian innovations; \emph{heavy tails}, Student-$t_3$
innovations of the same variance built from the same normal draws (negative binomial sizes on D8);
\emph{outliers}, each context point shifted with probability 0.03 by four to six robust standard deviations
(a demand multiplied by five on D8), with the held-out values left clean; and \emph{tested seasonality}, in
which the classical methods use $m = 12$ only if the 90\% autocorrelation test of the M4 benchmarks
\citep{makridakis2020m4} finds seasonality, applied when at least three cycles are observed, and $m = 1$
otherwise. TimesFM-3 never receives a period, so its forecasts are unchanged in the last version. A fifth
version, added later, applies the heavy tails and the outliers together, with the same draws.

\begin{table}[!htbp]
\centering
\begin{threeparttable}
\caption{Headline conclusions under randomised parameters and departures from the assumptions.}
\label{tab:robust-design}
\small\setlength{\tabcolsep}{4pt}
\begin{tabular}{lccccc}
\toprule
 & Random & Heavy tails & Outliers & Both & Tested seas. \\
\midrule
TimesFM-3 scenario wins (of 36) & 14 & 15 & 16 & 19 & 16 \\
TimesFM-3 median ratio to scenario-best & 1.029 & 1.038 & 1.004 & 1.000 & 1.017 \\
TimesFM-3 worst ratio to scenario-best & 1.56 & 1.54 & 1.49 & 1.56 & 1.55 \\
Smallest worst ratio, classical & 2.12 & 2.32 & 2.09 & 2.30 & 1.58 \\
Significant comparisons, won--lost & 93--29 & 100--28 & 101--27 & 104--24 & 100--27 \\
D4, $n \geq 48$: AutoARIMA better (of 3) & 3 & 3 & 2 & 2 & 3 \\
D8: Croston better on RMSSE (of 4) & 4 & 4 & 4 & 4 & 4 \\
D5, $n = 24$: AutoETS / seasonal naive ($m = 12$) & 2.03 & 2.05 & 1.91 & 1.95 & 2.25 \\
\bottomrule
\end{tabular}
\begin{tablenotes}[flushleft]\footnotesize
\item \textit{Note:} 200 replications per (process, length) scenario, 7\,200 series per column; every series draws its own parameters from the ranges in Table~S8 of the Supporting Information. Ratios use mean MASE over $h = 1, \dots, 12$. Significant comparisons are paired Wilcoxon tests of TimesFM-3 against each of the five general-purpose classical methods in each scenario (180 per column), Benjamini--Hochberg at 0.05 within each (column, opponent) family of 36 scenarios (the main design's families also include the three horizon slices). The D8 row compares TimesFM-3 with the best of Croston-SBA, TSB and ADIDA; the last row divides by seasonal naive with the true period in every column. ``Both'' combines the heavy tails and outliers of the two preceding columns (added later, same draws). In the tested-seasonality column the classical methods use $m = 12$ only where the M4 seasonality test finds seasonality; TimesFM-3's forecasts are those of the first column.
\end{tablenotes}
\end{threeparttable}
\end{table}

In the four versions with a known period the conclusions were qualitatively consistent with the main
design (\tableref{tab:robust-design}), although TimesFM-3's worst ratio lies outside the main design's
bootstrap interval: it is the best method in 14 to 19 scenarios, within 4\% of the best in the median scenario, and
its worst ratio rises to 1.49 to 1.56 (bootstrap intervals up to about 1.83), always on the logistic process
D7 at $n = 96$, while the smallest classical worst ratio is 2.09 to 2.32. TimesFM-3 wins 93 to 104 of 180 comparisons and loses 24 to 29, AutoARIMA again being the opponent it does not clearly beat. AutoARIMA's
advantage on D4 with at least four cycles persists (median over the three lengths 21\% to 26\%; with outliers it is ahead at two of the three lengths, by 15\% and 5\%, and with outliers and heavy
tails together by 16\% and 6\%), and the Croston family keeps the lower
RMSSE on D8 in every version.

\textbf{When the classical methods test for seasonality, the gap in worst ratios closes, but not because they
forecast better}: their smallest worst ratio is 1.58 (1.48 to 1.67, AutoARIMA), level with TimesFM-3's 1.55
(1.41 to 1.71). AutoARIMA's worst scenario is still D4 at $n = 24$, and its mean MASE there is unchanged (2.30);
what changes is the yardstick. At $n = 24$ the test cannot be applied, every series is treated as
non-seasonal, and seasonal naive loses the true period that made it the best method in that scenario (mean MASE
4.26 instead of 1.09), so TimesFM-3 (1.45) becomes the scenario-best. Much of the classical deficit in worst
ratios is thus measured against seasonal naive with a known period.

A response-surface regression of the per-series $\log_2$ ratio of TimesFM-3's MASE to AutoARIMA's on the
standardised parameters, $\log_2 n$ and the version, with standard errors clustered by parameter draw
(Supporting Information, Tables~S4 and~S5, and Figure~S2), shows that the dependence on the settings is
modest. The largest systematic effect is length: each doubling of $n$ moves the ratio against TimesFM-3 by about 18\% on D4 and 10\% on D5, and in its favour
on D6 and D8. Outliers, alone or with heavy tails, favour TimesFM-3 on the seasonal processes (14\% and 10\%;
14\% and 8\%). Among the
parameters only the demand probability of D8 has a large effect; the ratio is flat in the autoregressive
coefficient of D1 and the GARCH persistence of D9, grows in favour of TimesFM-3 with the size of the break in
D6, and varies with the inflection point of D7 in a way a linear term does not capture. The regressions
explain little of the per-series variation ($R^2 \leq 0.23$).

\textbf{Familiarity.} Because TimesFM-3 is pre-trained partly on synthetic series, its advantage could
reflect familiarity with some process families rather than their difficulty. In the randomised design with
Gaussian innovations, at matched difficulty (spectral entropy and length), its advantage over AutoARIMA is
about 4\% to 6\% larger on the processes
that no classical family specifies (D6--D8) than on D1--D5 (shift in the $\log_2$ ratio $-0.096$, 95\%
interval $-0.128$ to $-0.064$; median regression $-0.066$), which is consistent with an effect of familiarity but does not establish one: on D6--D8 AutoARIMA is also
misspecified by construction, which predicts the same shift.

\subsection{A longer horizon}
\label{sec:horizon48}

Calibration of foundation models is reported to degrade with the horizon \citep{adler2026calibrated}, and the
main design stops at 12 steps. The nine processes were therefore regenerated with the same seeds for a
48-step horizon and forecast by the six pre-registered methods, TimesFM-2.5 and Chronos-2 (Supporting
Information, Section~S7, Table~S14 and Figure~S3). Because the inflection of D7 sits at $0.6(n + H)$, the
longer horizon moves it into the context for $n = 200$, so that the series has already flattened when the
forecast starts. Over the first 12 steps the picture of \sectionref{sec:results} is recovered: without D7,
TimesFM-3's worst ratio to the best method is 1.31 and AutoARIMA's 2.11; over steps 13 to 24 TimesFM-3 still
has the smaller worst case (1.39 against 1.60). Over steps 25 to 48 it does not: \textbf{AutoARIMA has the
smaller worst case} (1.41 against 1.80, D7 excluded), although TimesFM-3 remains closer to the best method in the median scenario
(1.064 against 1.092, D7 excluded). On D7 at $n = 200$ all three foundation models forecast a decline while the
process stays at its plateau near 200: TimesFM-3's mean forecast falls to 128 by step 48, TimesFM-2.5's to 98
and Chronos-2's to 146, and their mean MASE over the 48 steps is 8.37, 12.22 and 5.30, against 1.11 for
seasonal naive and 2.51 for AutoARIMA. What in the models produces the decline cannot be established here.
On these regenerated series the coverage of TimesFM-3's 80\% intervals falls from 0.788 over steps 1 to 12 to
0.760 over steps 25 to 48, and AutoARIMA's from 0.760 to 0.720, so both drift away from nominal, as
\citet{adler2026calibrated} report for foundation models. The robustness of \sectionref{sec:estimability} is
thus a property of horizons up to about two years of monthly steps.

\section{Application to real data}
\label{sec:realdata}

Whether the foundation models have seen the M4 data cannot be settled. TimesFM-3's documented real-world
corpus, the GIFT-Eval pre-training collection with Wikipedia page views and Google Trends, is built to
exclude the GIFT-Eval test datasets, of which M4 Monthly is one \citep{timesfm3card, aksu2024gifteval,
giftevalpretrain}, but the exclusion works at the level of datasets and the synthetic part is undocumented;
TimesFM-2.5 lists the same sources \citep{timesfm25card}. The original Chronos models were trained on M4
Monthly with its final 18 observations, the official test period, held out \citep[Table~2]{ansari2024chronos},
so Chronos-Bolt's M4 forecasts may be in-domain rather than zero-shot; the corpora of Chronos-2 and TiRex were
not checked. The M4 tier is therefore secondary evidence; its official-test and truncation analyses are post hoc, and
\sectionref{sec:postcutoff} adds, post hoc, data observed after the documented training corpora.

\subsection{Representativeness of the simulated series}
\label{sec:representativeness}

Following \citet{kang2017instance}, each series is summarised by four features: normalised spectral
entropy (forecastability), STL trend strength, STL seasonal strength (period 12) and the first
autocorrelation of the first differences. The simulated series at $n = 96$ and 1\,000 M4 Monthly series
(\sectionref{sec:m4results}), each cut to the last 96 observations of its training period, are placed in
this space with the features standardised on the M4 series. An M4 series is \emph{covered} when some
simulated series lies at least as close to it as the 95th percentile of the distance from an M4 series to
its nearest other M4 series.

\begin{figure}[!htbp]
\centering
\includegraphics[width=\textwidth]{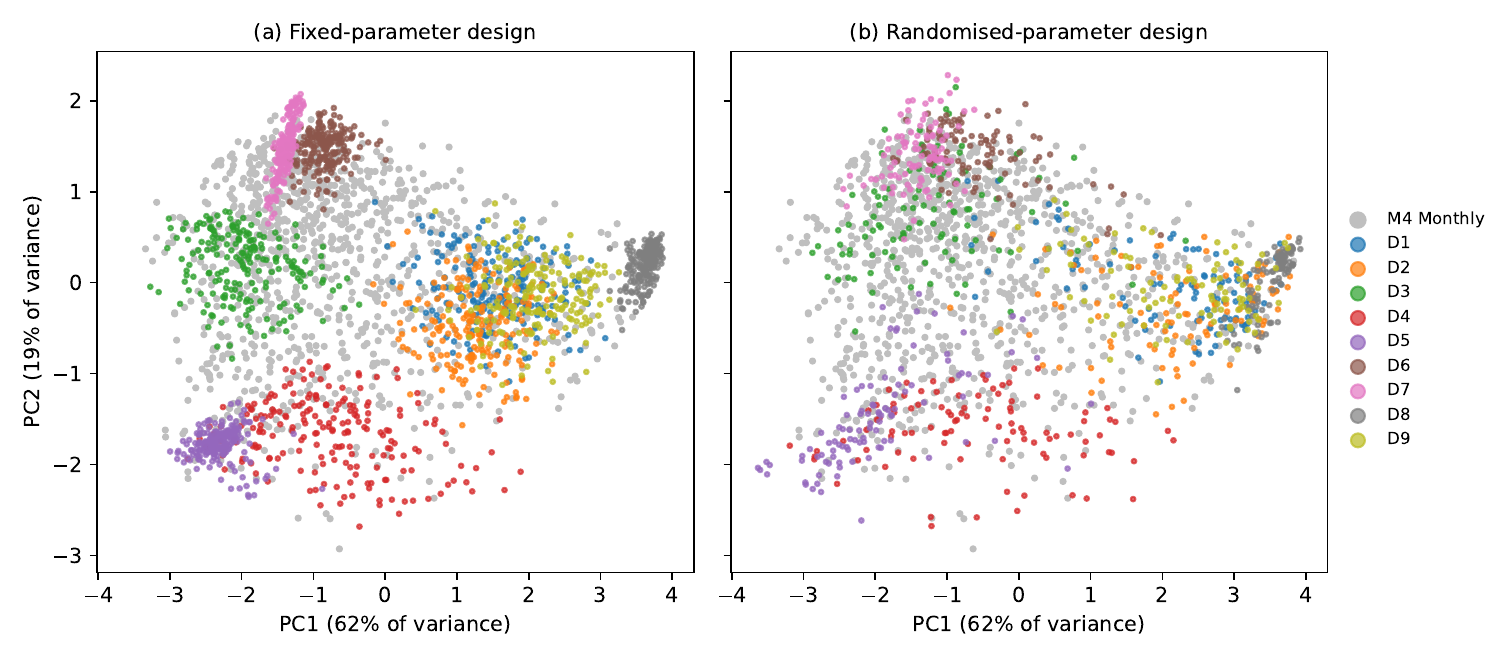}
\caption{Simulated and M4 Monthly series in a four-feature instance space (spectral entropy,
trend strength, seasonal strength and first-order autocorrelation of the differences), first two
principal components. Grey points are M4 series; coloured points are simulated series at
$n = 96$. (a) The fixed-parameter design covers 48.5\% of the M4 series; (b) the randomised-parameter
design covers 75.8\%, with coverage defined in \sectionref{sec:representativeness}.}
\label{fig:instance}
\end{figure}

The fixed design covers 48.5\% of the M4 series (bootstrap interval 45.2\% to 51.4\%;
\figureref{fig:instance}): each process forms a compact cluster and the regions between them, which hold
many real series, are empty. The randomised design of \sectionref{sec:departures} covers 75.8\% (73.2\% to
78.4\%). Across thresholds from the 80th to the 99th percentile, lengths of 48, 96 and 200, and a six-feature
space adding the first autocorrelation and lumpiness, the randomised design covers 76\% to 89\% of M4 at the
95th percentile and never less than the fixed design, also when the latter is subsampled to equal size;
conversely, 60\% to 92\% of the randomised series lie within the same distance of some M4 series, fewest at
$n = 200$. The parameter ranges were set before this statistic was first computed. M4 series are more
strongly trending than the simulated ones (median trend strength 0.87 against 0.63), and only 1.5\% lie
nearest to D8, because M4 has no intermittent series; the intermittent-demand results rest on the simulation
alone. Coverage of the feature space does not mean that results transfer series by series: predicting each
M4 series' $\log_2$ ratio of TimesFM-3's MASE to AutoARIMA's from its ten nearest simulated series gives a
Spearman correlation of 0.06 (95\% interval $-0.005$ to 0.125) with the realised ratio, no better than a
constant prediction.

\subsection{Rolling-origin evaluation}
\label{sec:m4results}

A seeded sample of 1\,000 series is drawn from the 32\,581 M4 Monthly series with at least 120 training
observations. Their training periods have a median of
281 observations (interquartile range 198 to 306), against 202 (82 to 306) for all 48\,000 series, so the
sample over-represents the long end of the population. Each series is evaluated at up to three origins 12 months apart with M4's horizon $h = 18$ (2\,932
windows), the latest being the official test period, forecast from the training series; consecutive windows overlap by six months, so the
errors of a series are averaged over its origins and the series are the independent units of the paired
Wilcoxon tests. TimesFM-3 (mean MASE 0.914) significantly outperforms seasonal naive (1.280; better on 80.1\%
of series), Theta (0.970; 56.8\%) and AutoETS (0.951; 54.1\%), and shows no detectable difference from
AutoARIMA (0.923; 52.6\%, adjusted $p = 0.27$) or the combination (0.907; 48.5\%, adjusted $p = 0.27$).

This ties TimesFM-3 with AutoARIMA on long, mostly seasonal series, rather than showing the AutoARIMA
advantage found on the simulated D4: real series mix trend, seasonality and irregular components in
proportions the simulation does not reproduce (\sectionref{sec:representativeness}), and an advantage from
contamination cannot be excluded either. The probabilistic results also diverge: TimesFM-3's 80\% intervals
cover 0.765 against 0.801 for the combination, yet its pinball loss is the lowest of the six methods (0.363,
against 0.365 for the combination and 0.374 for AutoARIMA)---a better distribution overall with too narrow an
80\% interval, the fragility of foundation-model calibration reported by \citet{adler2026calibrated}.

\subsection{The official M4 test period}
\label{sec:m4official}

To place the results against the published M4 results, the official 18-month test period, the latest of
the three origins above, was scored as in M4, with the overall weighted average (OWA) of sMAPE and MASE relative to the official Naive2 forecasts
\citep{makridakis2020m4}. The comparison includes the methods of \sectionref{sec:benchmarks} and the
published point forecasts of the ten best M4 submissions, among them the winning hybrid of exponential
smoothing and a recurrent network \citep{smyl2020hybrid} and the feature-based combination FFORMA
\citep{monteromanso2020fforma}; scored on all 48\,000 monthly series, the downloaded files reproduce the
published M4 results.

\begin{table}[!htbp]
\centering
\begin{threeparttable}
\caption{M4 Monthly official test period: accuracy, OWA and multiple comparisons.}
\label{tab:m4official}
\small\setlength{\tabcolsep}{4pt}
\begin{tabular}{lcccccc}
\toprule
Method & sMAPE & MASE & OWA & Coverage & MSIS & Mean rank \\
\midrule
FFORMA (M4, 2nd) & 7.82 & 0.889 & 0.831 & -- & -- & 9.85$^\ast$ \\
Jaganathan \& Prakash (M4, 4th) & 7.88 & 0.902 & 0.841 & -- & -- & 10.06$^\ast$ \\
ES-RNN, Smyl (M4, 1st) & 7.98 & 0.907 & 0.848 & -- & -- & 10.72$^\ast$ \\
Pawlikowski et al.\ (M4, 3rd) & 7.95 & 0.912 & 0.849 & -- & -- & 10.38$^\ast$ \\
TiRex & 8.04 & 0.916 & 0.855 & 0.801 & 4.40 & 10.77$^\ast$ \\
TimesFM-3, evaluator settings & 8.01 & 0.923 & 0.857 & 0.765 & 4.50 & -- \\
Petropoulos \& Svetunkov (M4, 6th) & 8.09 & 0.916 & 0.858 & -- & -- & 10.46$^\ast$ \\
Chronos-2 & 8.11 & 0.917 & 0.860 & 0.783 & 4.47 & 10.69$^\ast$ \\
Fiorucci \& Louzada (M4, 5th) & 8.09 & 0.920 & 0.860 & -- & -- & 10.50$^\ast$ \\
TimesFM-3 & 8.05 & 0.929 & 0.862 & 0.761 & 4.54 & 10.69$^\ast$ \\
TimesFM-2.5 & 8.15 & 0.922 & 0.864 & 0.812 & 4.53 & 11.03 \\
Shaub (M4, 7th) & 8.06 & 0.933 & 0.864 & -- & -- & 11.01 \\
Combination (ETS, ARIMA, DOTM) & 8.16 & 0.928 & 0.867 & 0.789 & 4.74 & -- \\
Combination (Theta, ETS, ARIMA) & 8.16 & 0.929 & 0.868 & 0.789 & 4.76 & 10.58$^\ast$ \\
TimesFM-2.5, settings off & 8.25 & 0.932 & 0.874 & 0.802 & 4.61 & -- \\
Doornik et al.\ (M4, 9th) & 8.21 & 0.945 & 0.878 & -- & -- & 11.64 \\
Pedregal et al.\ (M4, 10th) & 8.31 & 0.953 & 0.887 & -- & -- & 11.88 \\
AutoARIMA & 8.50 & 0.939 & 0.891 & 0.779 & 4.91 & 11.13 \\
Chronos-Bolt & 8.40 & 0.971 & 0.900 & 0.800 & 4.70 & 11.91 \\
AutoETS & 8.45 & 0.966 & 0.900 & 0.784 & 4.98 & 11.82 \\
M4 Comb & 8.50 & 0.982 & 0.911 & -- & -- & 12.65 \\
Legaki \& Koutsouri (M4, 8th) & 8.51 & 0.995 & 0.917 & -- & -- & 12.76 \\
DOTM & 8.59 & 0.996 & 0.922 & 0.728 & 5.27 & 13.03 \\
Theta & 8.59 & 1.006 & 0.926 & 0.723 & 5.36 & 13.00 \\
Naive2 & 9.22 & 1.092 & 1.000 & -- & -- & -- \\
Seasonal naive & 10.39 & 1.306 & 1.161 & 0.780 & 6.56 & 16.44 \\
\bottomrule
\end{tabular}
\begin{tablenotes}[flushleft]\footnotesize
\item \textit{Note:} 1\,000 M4 Monthly series, each forecast for the official 18-month test period from its full training history. sMAPE and MASE as defined in M4; OWA relative to the official Naive2 forecasts. Coverage and MSIS use the nominal 80\% interval, the widest that every probabilistic method can form from nine deciles. Mean rank: average rank of per-series MASE among 22 methods, one configuration per model (Friedman $\chi^2 = 1054$, $p < 10^{-100}$); $^\ast$ marks methods whose mean rank is within the Nemenyi critical difference (1.04) of the best; -- marks alternative configurations and Naive2, left out of the ranking. Published M4 submissions, ranked by their overall M4 position, provide point forecasts only \citep{makridakis2020m4, smyl2020hybrid, monteromanso2020fforma}.
\end{tablenotes}
\end{threeparttable}
\end{table}

Theta and the Comb benchmark improve on Naive2 by 7\% to 9\% in OWA (\tableref{tab:m4official}), in line
with the competition. \textbf{Four M4 entries are ahead of every foundation model}: FFORMA (OWA 0.831),
Jaganathan and Prakash (0.841), Smyl's hybrid (0.848) and Pawlikowski et al.\ (0.849). TimesFM-3 (0.862;
0.857 with its evaluator settings) is level with the entries ranked fifth to seventh in M4 (0.858 to 0.864),
TiRex (0.855) and Chronos-2 (0.860) are close to it, and the classical combination (0.868) and AutoARIMA
(0.891) follow. On this sample the M4 entries' OWA differs from their published monthly values by up to
0.03, as much as the gaps between them. A Friedman test on per-series ranks, with one configuration per model
and without Naive2 \citep{benavoli2016should}, rejects the equality of the 22 methods, but ten lie within the
Nemenyi critical difference of the best mean rank \citep{demsar2006statistical}: six M4 entries, the
classical combination, TimesFM-3, Chronos-2 and TiRex. On real monthly series with long histories, zero-shot
TimesFM-3 matches good specialised methods of 2018 without per-series tuning, but not the best of them.
Reweighting the sample to the length distribution of the population it was drawn from leaves TimesFM-3's
OWA at 0.861 and the four leading entries unchanged; the shorter third of the M4 Monthly series is not
represented, and the truncation experiment below stands in for it.

\subsection{Truncated histories}
\label{sec:truncation}

The official test period was also forecast from only the last 24, 48 or 96 observations of each training
series, with the full-history MASE denominator so that only the information available to the methods
changes (\tableref{tab:truncation}). The simulation's estimability result reappears: with 24 observations,
where the classical methods again have two cycles for a period of 12, the foundation models lead---TiRex
1.027, TimesFM-3 1.046, TimesFM-2.5 1.050---against 1.096 for the best classical method, the combination,
and 1.175 for AutoARIMA; the difference was not tested. With 48 observations the combination is level with
them (0.955 against 0.952 to 0.972), and with the full history the leading methods are within 2\% of one
another. Truncation also moves the information set away from the full series a model could have been trained
on, although it cannot rule out that the series were seen.

\begin{table}[!htbp]
\centering
\begin{threeparttable}
\caption{M4 Monthly with truncated contexts: mean MASE by context length.}
\label{tab:truncation}
\small\setlength{\tabcolsep}{4pt}
\begin{tabular}{lcccc}
\toprule
Method & Last 24 & Last 48 & Last 96 & Full \\
\midrule
TimesFM-3 & 1.046 & 0.961 & 0.942 & 0.929 \\
TimesFM-3, evaluator settings & 1.045 & 0.960 & 0.936 & 0.923 \\
TimesFM-2.5 & 1.050 & 0.972 & 0.952 & 0.922 \\
Chronos-Bolt & 1.158 & 1.023 & 0.965 & 0.971 \\
Chronos-2 & 1.087 & 0.952 & 0.935 & 0.917 \\
TiRex & 1.027 & 0.956 & 0.949 & 0.916 \\
AutoARIMA & 1.175 & 1.019 & 0.979 & 0.939 \\
AutoETS & 1.129 & 0.996 & 0.995 & 0.966 \\
Theta & 1.150 & 1.011 & 1.003 & 1.006 \\
DOTM & 1.167 & 1.012 & 0.991 & 0.996 \\
Combination (Theta, ETS, ARIMA) & 1.096 & 0.955 & 0.947 & 0.929 \\
Combination (ETS, ARIMA, DOTM) & 1.111 & 0.959 & 0.945 & 0.928 \\
Seasonal naive & 1.306 & 1.306 & 1.306 & 1.306 \\
\bottomrule
\end{tabular}
\begin{tablenotes}[flushleft]\footnotesize
\item \textit{Note:} Mean MASE on the M4 official test period when each method sees only the last 24, 48 or 96 training observations, or the full history (median 281 observations). All columns use the full-history MASE denominator, so only the information given to the methods changes. Seasonal naive uses only the last 12 observations and is unaffected.
\end{tablenotes}
\end{threeparttable}
\end{table}

\subsection{Data observed after the training corpora}
\label{sec:postcutoff}

Neither the simulation nor M4 can show how the foundation models behave on real data they cannot have
memorised. The monthly FRED-MD database \citep{mccracken2016fredmd}, in its August 2026 vintage, offers such
data: 101 of its 126 series are complete from 1990 to June 2026, and every method forecast January 2025 to
June 2026 (18 steps) from the 240 months to December 2024, raw levels without transformation. The evaluation
window post-dates the documented corpora of TimesFM-3 and TimesFM-2.5 (to November 2023) and the releases of
Chronos-Bolt; TiRex and Chronos-2 were released within it, and restricting the window to November 2025
onwards leaves AutoARIMA first (mean MASE 0.722) and puts TimesFM-3 behind every method except the naive
benchmarks (0.857). On this tier \textbf{TimesFM-3 is statistically level with the classical methods}
(Supporting Information, Table~S16): AutoARIMA has the lowest mean MASE (0.540), TiRex and the combination
follow (0.556), and TimesFM-3 has 0.607, with a median (0.426) close to AutoARIMA's (0.411); paired Wilcoxon
tests with Benjamini--Hochberg correction separate TimesFM-3 only from seasonal naive, which it beats, and a
Friedman test puts every method except seasonal naive within the Nemenyi critical difference of the best
mean rank.
TimesFM-3's mean is raised by a few series on which it extended a recent movement far beyond what followed,
such as the 2024 cuts in short-term interest rates. The tier is small and dominated by trending
macroeconomic levels, and it rules out memorised test values, not familiarity with the series' past.

\section{Discussion}
\label{sec:discussion}

\subsection{Summary of findings}
\label{sec:findings}

\tableref{tab:findings} collects the findings with their evidence, their status with respect to the protocol
and the conditions under which they hold.

\begin{table}[!htbp]
\centering
\begin{threeparttable}
\caption{Summary of findings.}
\label{tab:findings}
\small
\begin{tabular}{>{\raggedright\arraybackslash}p{5.6cm}>{\raggedright\arraybackslash}p{2.1cm}>{\raggedright\arraybackslash}p{2.4cm}>{\raggedright\arraybackslash}p{3.9cm}}
\toprule
Finding & Evidence & Status & Holds within \\
\midrule
Neither family dominates; AutoARIMA is the only classical method TimesFM-3 does not clearly beat &
  \sectionref{sec:overall}, \figureref{fig:forest} & Pre-registered & Nine processes, $H = 12$ \\[2pt]
TimesFM-3's worst ratio to the best method is 1.31, every classical method's at least 2.2 &
  \tableref{tab:robust}, \sectionref{sec:mcse} & Post-hoc summary & $H = 12$; relative to the methods compared \\[2pt]
The largest failures of automatic ARIMA and ETS come from seasonal models that two cycles cannot estimate &
  \sectionref{sec:estimability}, \sectionref{sec:checks} & Pre-registered, checked post hoc & $m = 12$, $n = 24$; reproduced in R \\[2pt]
With four or more cycles AutoARIMA is 21--24\% more accurate on seasonal ARIMA data &
  \sectionref{sec:regimes} & Pre-registered & D4, D5 \\[2pt]
The robustness is specific to TimesFM-3 among five foundation models &
  \sectionref{sec:benchmarks}, \figureref{fig:scatter} & Post hoc & $H = 12$ \\[2pt]
On intermittent demand no advantage over simple history-based benchmarks &
  \sectionref{sec:d8-extended} & Post hoc & An independent process \\[2pt]
At steps 25--48 AutoARIMA has the smaller worst case; the three foundation models run there forecast a decline on a plateau &
  \sectionref{sec:horizon48} & Post hoc & Regenerated design \\[2pt]
On M4, TimesFM-3 is level with the fifth- to seventh-ranked entries, behind the four best &
  \sectionref{sec:m4official} & Post hoc & 1\,000 long monthly series \\[2pt]
On data observed after the documented training corpora, TimesFM-3 is level with the classical methods &
  \sectionref{sec:postcutoff} & Post hoc & 101 FRED-MD series, 2025--2026 \\
\bottomrule
\end{tabular}
\end{threeparttable}
\end{table}

\subsection{Robustness versus peak accuracy}
\label{sec:tradeoff}

The classical methods' 20 scenario wins are spread over five methods, so which classical method wins is not
knowable in advance without the model selection a foundation model is meant to remove. TimesFM-3, one
untuned model, is not significantly worse than the best classical method in 25 of the 36 scenarios; the best
classical method is ahead in the other 11, over five processes, by small margins on D2 and D3 and by large
ones only on D4, D5 and D7 at $n = 96$ (these comparisons with a selected opponent are descriptive). The
evidence therefore supports neither the conclusion that foundation models have superseded classical methods
nor the opposite, but a narrower one: \textbf{in this design TimesFM-3 behaved robustly rather than with peak
accuracy}. It is rarely the best method and never among the worst, whereas the automatic classical methods
are sometimes excellent and sometimes catastrophic. For one well-understood series that trade is
unattractive; for many heterogeneous series and no capacity to diagnose each one, it favoured TimesFM-3 here,
within the horizons, processes and yardstick set out in \tableref{tab:findings}.

\subsection{Practical considerations}
\label{sec:practical}

The considerations below are conditional on the processes, lengths and horizons examined. They describe how
the methods behaved here and are offered as diagnostic starting points to be checked on the user's own
data, not as rules.

\textbf{Many heterogeneous series without per-series tuning.} This is where TimesFM-3's robustness mattered
most: it was never far from the best method of a scenario. Where a model can be identified for a single series,
TimesFM-3 gains little over AutoARIMA, which keeps parameters, diagnostics and interval theory that
can be explained and audited; when TimesFM-3 forecasts poorly, there is nothing to inspect, and its quantiles
are learned outputs with no accompanying theory, well calibrated on average in the simulation but not on M4.
In regulated settings where a forecast must be justified, these differences may outweigh the accuracy
comparison.

\textbf{Seasonal series.} With at least four observed cycles AutoARIMA was clearly ahead on seasonal ARIMA
data, and seasonal naive was also ahead of TimesFM-3 there. With two cycles the automatic classical methods
fell back to non-seasonal models and failed, and seasonal naive, or TimesFM-3, was the safer choice.

\textbf{Intermittent demand.} Simple benchmarks built from the history were as good as anything tested: the
empirical deciles of the history gave the lowest pinball loss, and its mean, like the Croston family, the
lowest RMSSE, while TimesFM-3's median forecast was no better than forecasting zero.

\textbf{Long horizons.} Beyond about two years of monthly steps AutoARIMA had the smaller worst case, and the three foundation models run there forecast a decline on a process that had levelled off, so long-horizon forecasts from
these models should be checked against the recent level before use.

\textbf{Cost.} Timed on the same 200 seasonal series of length 96, after a warm-up run, seasonal naive
took 2.8\,ms per series, Theta 9.1\,ms, AutoETS 106.7\,ms and AutoARIMA 2\,369.7\,ms, each on one CPU core,
and TimesFM-3 24.9\,ms batched on one consumer GPU and 47.1\,ms on the CPU (12 threads), excluding a
one-off model load of 2 to 4\,s, so TimesFM-3 does not require a GPU. Without the warm-up, which leaves in
the one-off compilation of the \pkg{StatsForecast} routines, the classical timings differ by less than
10\%. In the main run the four
classical methods took 7.3 minutes on 22 cores against 3.2 minutes for TimesFM-3, a ratio of $2.3\times$ that
is dominated by AutoARIMA and depends on the hardware. The classical methods hold a handful of parameters per
series; TimesFM-3 keeps a checkpoint of about 1.3\,GB resident, whatever the number of series.

\textbf{Licence.} TimesFM-3's weights are released under the TimesFM Non-Commercial License v1.0, which
grants use ``solely for your Non-Commercial Purposes'' and requires a separate licence for any ``commercial
or production activity'' \citep{timesfm3repo, timesfm3card} (accessed 23 September 2026). The source code is
Apache-2.0, as are the weights of TimesFM-2.5, Chronos-Bolt and Chronos-2, but the version-3 weights are not:
\textbf{for a commercial application, TimesFM-3 is not licensed as released}, whatever its accuracy, and the
alternatives are TimesFM-2.5 (less robust in the simulation, level on M4), separate terms or a classical
method. Licensing is routinely omitted from model comparisons although it decides which models an entire
class of users may use.

\subsection{Threats to validity}
\label{sec:limitations}

\emph{Construct.} The summaries that carry the robustness finding---worst ratios and win counts---were
defined after the protocol and depend on the methods and scenarios compared, notably on seasonal naive with the
known period; on D8 the verdict depends on the error measure. \emph{Internal.} No method was tuned by hand,
and the classical methods were run in \pkg{StatsForecast} at its defaults; the cross-check with \pkg{forecast}
reproduced the failures but found one implementation failure, and an expert who identified each process
could select better models. The protocol was frozen before any forecast but held in the replication archive
rather than deposited with an external registry, so its only external timestamp postdates the forecasts.
Everything in \sectionref{sec:posthoc} and \sectionref{sec:realdata} except the rolling-origin evaluation is
post hoc. Only the data of \sectionref{sec:postcutoff} are known to post-date the documented training
corpora. \emph{External.} Simulated
series are clean and free of the measurement error, calendar effects and regime changes of applied work; the
M4 tier addresses this only in part, and the randomised design still leaves about a quarter of M4 monthly
series, and all other frequencies, outside the regions studied. The nine process families are fixed, the
seasonal period is 12, the pre-registered horizon is 12 steps and the longer one was examined in one
secondary check, and the departures from the assumptions were applied one at a time and in one combination. TimesFM-3 was used
zero-shot and univariate, leaving its multivariate and covariate capabilities untested; fine-tuning on one
series of 24 to 200 points is not meaningful for a model of this size, and fine-tuning on the DGPs would remove
the separation between model and test data. Five foundation models were evaluated and they behave
differently, so conclusions about one do not transfer to the class; other families, such as Moirai
\citep{woo2024moirai, liu2025moirai2}, Sundial, Time-MoE and Toto, were not included.

\section{Conclusion}
\label{sec:conclusion}

Under a pre-registered protocol and on data no model could have seen, TimesFM-3 and classical methods were
compared on 7\,200 series from nine known processes. Neither family dominates: TimesFM-3 wins most
significant comparisons, but not against AutoARIMA, and with at least four seasonal cycles AutoARIMA is
clearly more accurate. What distinguishes TimesFM-3 is its worst case, an error never more than 1.31 times that of the best method in a scenario against at least
2.2 times for every classical method, and among five foundation models it is
the only one with this property. The property is bounded: it is measured relative to the methods compared,
it holds for horizons up to about two years of monthly steps, and on intermittent demand simple benchmarks
built from the history do as well. On the M4 test period TimesFM-3 is level with good but not the best M4 entries, and on macroeconomic data
observed after its documented training data it is level with the classical methods. The finding most likely to transfer concerns the classical side, where the mechanism can be
examined: what matters is whether a classical model is \emph{estimable} from the data, not only whether it is
correctly \emph{specified}.

The study is exploratory. It describes how one black-box model behaves on data of known structure, does not
explain that behaviour, and remains conditional on the processes, parameter ranges, horizons and benchmarks
examined. For commercial users the non-commercial licence of TimesFM-3's weights decides the question before
accuracy does.

\section{Computational details and reproducibility}
\label{sec:repro}

All results were produced with \proglang{Python}~3.13 \citep{python}, \pkg{statsforecast}~2.1.1
\citep{statsforecast}, \pkg{timesfm}~3.0.1 with the \path{google/timesfm-3.0-pytorch} weights (revision
\code{43046b8}) \citep{timesfm3repo}, \pkg{chronos-forecasting}~2.3.2 with \path{amazon/chronos-bolt-base}
(revision \code{5d9f166}) and \path{amazon/chronos-2} (revision \code{29ec376}), \pkg{tirex-ts}~1.4.2 with
\path{NX-AI/TiRex} (revision \code{63c7409}), the \path{google/timesfm-2.5-200m-pytorch} weights (revision
\code{1d95242}), \proglang{R}~4.4.1 with \pkg{forecast}~9.0.2 and \pkg{smooth}~4.5.2 for the cross-check and iETS,
\pkg{PyTorch}~2.6.0 with CUDA~12.4 and cuDNN~9.1 \citep{paszke2019pytorch}, \pkg{NumPy}
\citep{harris2020numpy}, \pkg{pandas} \citep{mckinney2010pandas}, \pkg{SciPy} \citep{virtanen2020scipy} and
\pkg{Matplotlib} \citep{hunter2007matplotlib}, on an NVIDIA GTX 1660 Ti (driver 591.86) with a 24-core CPU
and 32\,GB of memory; parallel classical runs use 22 worker processes. The full package list is pinned in the
replication archive. Every simulated series is seeded from its own identifiers, so the study regenerates
exactly, independently of core count or execution order, and each cached stage can be rerun alone. The
foundation models were run in 32-bit arithmetic without random sampling; rerunning two TimesFM-3 scenarios
reproduced the stored quantiles exactly, although other hardware may differ in the last digits because GPU
floating-point reductions are not associative. The protocol, with its five hypotheses, was last modified on 8
September 2026 at 20:01, 22 minutes before the first forecast file was written (local file-system
timestamps); its first external timestamp is the public archive of 9 September 2026, which postdates the
forecasts, and it has not been edited since. The archive holds a dated log of the fifteen departures from
it, each reasoned, and every post-hoc analysis is marked as such. Every DOI in the
bibliography was resolved and compared with its entry.

\section*{Acknowledgements}
First and foremost, the authors thank Allah, the Almighty, for everything He has given them, including
the ability to complete this work. The authors used a large language model to improve the language and
readability of this manuscript and to assist with the implementation of the simulation code; they
reviewed and edited all content and take full responsibility for it.

\section*{Funding}
No funding was received for this work.

\section*{Conflict of interest}
The authors declare no conflicts of interest.

\section*{Ethics approval}
Not applicable; the study uses simulated data and publicly available data (M4 competition data and
forecasts, FRED-MD), and
involves no human participants. Permission to reproduce material from other sources: not applicable.

\section*{Data availability statement}
The complete research compendium is archived on Zenodo, \url{https://doi.org/10.5281/zenodo.22681072} (concept DOI, resolving to all versions), and mirrored at \url{https://github.com/ebrahimkhaled/timesfm-classical-paper}: the pre-registered protocol, the deviation log, the entire pipeline, the per-series metrics, the raw forecasts and quantiles, every figure and the manuscript source. The simulation data are generated by that code from documented seeds; the real-data tier uses the publicly available M4 competition data \citep{makridakis2020m4}, which the code re-fetches, with the seeded 1\,000-series sample included, and the FRED-MD database (August 2026 vintage)
\citep{mccracken2016fredmd} and the published M4 forecasts, which are publicly available and are fetched
by the code rather than archived. The raw forecasts are deposited so that they can be re-scored under other losses. TimesFM-3's weights are not redistributed; the code downloads them from Hugging Face.

\bibliographystyle{plainnat}
\bibliography{refs}

\clearpage
\appendix
\setcounter{section}{0}\setcounter{table}{0}\setcounter{figure}{0}
\renewcommand{\thesection}{S\arabic{section}}
\renewcommand{\thetable}{S\arabic{table}}
\renewcommand{\thefigure}{S\arabic{figure}}
\begin{center}\Large\bfseries Supporting Information\end{center}

\section{Results under the mean absolute percentage error}
\label{si:mape}

\tableref{tab:mape} reports mean MAPE by process, alongside the percentage of series on which
it cannot be computed. The D8 row illustrates the problem: every intermittent series contains a zero,
so MAPE is undefined for all of them, and the process on which the error measures disagree most drops out of the comparison.

The rankings also move. On D4, MAPE places seasonal naive first (2.360) and TimesFM-3 second
(2.801), whereas MASE places AutoARIMA first at every length from $n = 48$. MAPE's asymmetry
between over- and under-forecasts is enough to reorder the methods on the same forecasts.

\begin{table}[!htbp]
\centering
\begin{threeparttable}
\caption{Mean MAPE over $h = 1,\dots,12$ and percentage of series on which MAPE is undefined.}
\label{tab:mape}
\small
\begin{tabular}{lrrrrrrr}
\toprule
DGP & \% undefined & Seas.\ naive & Theta & AutoETS & AutoARIMA & Combin. & \textbf{TimesFM-3} \\
\midrule
D1 & 0\% & 2.785 & 2.737 & 2.681 & 2.336 & 2.501 & \textbf{2.283} \\
D2 & 0\% & 3.291 & 3.341 & 3.277 & 2.696 & 2.971 & \textbf{2.641} \\
D3 & 0\% & 9.509 & 9.278 & 9.875 & 10.078 & 9.393 & \textbf{9.214} \\
D4 & 0\% & \textbf{2.360} & 7.471 & 4.986 & 2.862 & 4.538 & 2.801 \\
D5 & 0\% & 2.971 & 3.583 & 3.270 & 2.436 & 2.916 & \textbf{2.256} \\
D6 & 0\% & 2.011 & 2.379 & 1.835 & 1.902 & 1.925 & \textbf{1.822} \\
D7 & 0\% & 7.634 & 4.631 & 2.979 & 2.538 & 2.767 & \textbf{2.184} \\
D8 & 100\% & --- & --- & --- & --- & --- & --- \\
D9 & 0\% & 1.861 & 1.769 & 1.741 & 1.470 & 1.599 & \textbf{1.457} \\
\bottomrule
\end{tabular}
\begin{tablenotes}[flushleft]\footnotesize
\item \textit{Note:} On the intermittent process D8 every series contains a zero, so MAPE cannot be computed for it. The best value in each row is in bold.
\end{tablenotes}
\end{threeparttable}
\end{table}

\section{Monte Carlo standard errors}
\label{si:mcse}

\tableref{tab:si-mcse} gives the mean MASE of the six methods of the main design in every scenario,
with its Monte Carlo standard error.

\begin{table}[!htbp]
\centering
\begin{threeparttable}
\caption{Mean MASE and Monte Carlo standard errors, main design.}
\label{tab:si-mcse}
\scriptsize\setlength{\tabcolsep}{3pt}\renewcommand{\arraystretch}{0.9}
\begin{tabular}{llcccccc}
\toprule
DGP & $n$ & Seasonal naive & Theta & AutoETS & AutoARIMA & Combination & TimesFM-3 \\
\midrule
D1 & 24 & 1.646 (0.056) & 1.682 (0.055) & 1.568 (0.050) & 1.485 (0.048) & 1.538 (0.048) & 1.417 (0.042) \\
D1 & 48 & 1.625 (0.058) & 1.586 (0.055) & 1.570 (0.054) & 1.415 (0.043) & 1.478 (0.048) & 1.381 (0.039) \\
D1 & 96 & 1.626 (0.054) & 1.572 (0.053) & 1.564 (0.053) & 1.324 (0.039) & 1.435 (0.044) & 1.307 (0.033) \\
D1 & 200 & 1.615 (0.061) & 1.576 (0.057) & 1.576 (0.057) & 1.266 (0.035) & 1.417 (0.046) & 1.254 (0.033) \\
D2 & 24 & 1.905 (0.066) & 2.003 (0.073) & 1.886 (0.069) & 1.708 (0.087) & 1.788 (0.065) & 1.675 (0.062) \\
D2 & 48 & 1.838 (0.068) & 1.860 (0.069) & 1.834 (0.068) & 1.575 (0.053) & 1.688 (0.058) & 1.518 (0.042) \\
D2 & 96 & 1.810 (0.057) & 1.812 (0.057) & 1.811 (0.057) & 1.517 (0.038) & 1.643 (0.045) & 1.455 (0.035) \\
D2 & 200 & 1.922 (0.067) & 1.919 (0.067) & 1.921 (0.067) & 1.374 (0.034) & 1.654 (0.050) & 1.383 (0.033) \\
D3 & 24 & 4.411 (0.213) & 4.483 (0.219) & 4.892 (0.235) & 5.275 (0.256) & 4.659 (0.224) & 4.656 (0.236) \\
D3 & 48 & 4.252 (0.208) & 4.179 (0.206) & 4.547 (0.207) & 4.679 (0.239) & 4.313 (0.203) & 4.274 (0.205) \\
D3 & 96 & 4.125 (0.191) & 4.135 (0.189) & 3.938 (0.179) & 3.940 (0.179) & 3.871 (0.172) & 4.023 (0.177) \\
D3 & 200 & 4.512 (0.210) & 4.427 (0.211) & 4.275 (0.199) & 4.195 (0.201) & 4.199 (0.199) & 4.297 (0.198) \\
D4 & 24 & 1.029 (0.026) & 4.146 (0.110) & 3.602 (0.101) & 2.284 (0.079) & 3.084 (0.072) & 1.321 (0.031) \\
D4 & 48 & 1.021 (0.024) & 1.944 (0.048) & 1.720 (0.032) & 1.014 (0.030) & 1.395 (0.028) & 1.295 (0.027) \\
D4 & 96 & 1.005 (0.024) & 2.910 (0.078) & 2.407 (0.055) & 0.886 (0.020) & 1.839 (0.040) & 1.124 (0.022) \\
D4 & 200 & 1.053 (0.022) & 4.155 (0.117) & 1.085 (0.024) & 0.886 (0.021) & 1.700 (0.043) & 1.161 (0.020) \\
D5 & 24 & 1.139 (0.038) & 3.154 (0.102) & 3.025 (0.093) & 1.574 (0.043) & 2.474 (0.069) & 1.152 (0.038) \\
D5 & 48 & 1.108 (0.032) & 0.956 (0.038) & 0.741 (0.027) & 0.717 (0.024) & 0.731 (0.023) & 0.815 (0.029) \\
D5 & 96 & 1.076 (0.033) & 0.757 (0.025) & 0.761 (0.029) & 0.732 (0.027) & 0.713 (0.027) & 0.761 (0.029) \\
D5 & 200 & 1.040 (0.030) & 0.759 (0.020) & 0.681 (0.024) & 0.626 (0.022) & 0.645 (0.022) & 0.689 (0.024) \\
D6 & 24 & 0.801 (0.025) & 1.327 (0.038) & 0.766 (0.024) & 0.802 (0.025) & 0.837 (0.025) & 0.807 (0.036) \\
D6 & 48 & 0.947 (0.028) & 1.115 (0.031) & 0.869 (0.023) & 0.925 (0.027) & 0.924 (0.025) & 0.825 (0.022) \\
D6 & 96 & 1.006 (0.028) & 0.963 (0.026) & 0.905 (0.023) & 0.920 (0.023) & 0.916 (0.023) & 0.865 (0.021) \\
D6 & 200 & 0.934 (0.023) & 0.840 (0.017) & 0.825 (0.017) & 0.833 (0.017) & 0.827 (0.017) & 0.844 (0.019) \\
D7 & 24 & 5.417 (0.080) & 3.179 (0.058) & 1.272 (0.042) & 1.083 (0.037) & 1.680 (0.048) & 1.142 (0.033) \\
D7 & 48 & 4.243 (0.067) & 2.262 (0.048) & 1.602 (0.061) & 1.076 (0.051) & 0.970 (0.038) & 0.751 (0.022) \\
D7 & 96 & 1.655 (0.042) & 0.821 (0.020) & 1.305 (0.047) & 1.329 (0.048) & 1.097 (0.036) & 0.959 (0.032) \\
D7 & 200 & 0.917 (0.026) & 1.301 (0.031) & 0.966 (0.028) & 0.994 (0.030) & 0.954 (0.027) & 0.933 (0.025) \\
D8 & 24 & 1.019 (0.054) & 0.999 (0.034) & 1.047 (0.058) & 0.996 (0.032) & 1.010 (0.039) & 0.780 (0.031) \\
D8 & 48 & 1.066 (0.066) & 0.984 (0.036) & 0.988 (0.033) & 0.962 (0.024) & 0.974 (0.028) & 0.708 (0.030) \\
D8 & 96 & 0.932 (0.047) & 0.923 (0.017) & 0.926 (0.015) & 0.924 (0.015) & 0.923 (0.015) & 0.683 (0.022) \\
D8 & 200 & 1.005 (0.052) & 0.943 (0.017) & 0.939 (0.015) & 0.939 (0.015) & 0.940 (0.015) & 0.694 (0.023) \\
D9 & 24 & 1.547 (0.061) & 1.518 (0.057) & 1.454 (0.057) & 1.372 (0.057) & 1.404 (0.054) & 1.321 (0.049) \\
D9 & 48 & 1.588 (0.056) & 1.486 (0.054) & 1.476 (0.053) & 1.244 (0.041) & 1.353 (0.046) & 1.239 (0.039) \\
D9 & 96 & 1.430 (0.051) & 1.342 (0.048) & 1.333 (0.048) & 1.098 (0.034) & 1.213 (0.040) & 1.104 (0.034) \\
D9 & 200 & 1.577 (0.056) & 1.503 (0.053) & 1.504 (0.053) & 1.190 (0.032) & 1.341 (0.042) & 1.200 (0.035) \\
\bottomrule
\end{tabular}
\begin{tablenotes}[flushleft]\footnotesize
\item \textit{Note:} Mean MASE over $h = 1, \dots, 12$ with its Monte Carlo standard error in parentheses (standard deviation over the 200 replications divided by $\sqrt{200}$).
\end{tablenotes}
\end{threeparttable}
\end{table}

\section{Coverage by process}
\label{si:coverage}

\tableref{tab:si-coverage} gives the empirical coverage of the nominal 80\% intervals of the four
foundation-model configurations and of AutoARIMA and AutoETS, by process.

\begin{table}[!htbp]
\centering
\begin{threeparttable}
\caption{Coverage of 80\% prediction intervals by process.}
\label{tab:si-coverage}
\small\setlength{\tabcolsep}{4pt}
\begin{tabular}{lcccccc}
\toprule
Process & TimesFM-3 & TimesFM-3 (eval.) & TimesFM-2.5 & Chronos-Bolt & AutoARIMA & AutoETS \\
\midrule
D1 & 0.817 & 0.819 & 0.787 & 0.814 & 0.777 & 0.852 \\
D2 & 0.817 & 0.818 & 0.782 & 0.824 & 0.766 & 0.874 \\
D3 & 0.704 & 0.707 & 0.703 & 0.721 & 0.706 & 0.618 \\
D4 & 0.846 & 0.856 & 0.810 & 0.769 & 0.756 & 0.661 \\
D5 & 0.818 & 0.830 & 0.799 & 0.833 & 0.724 & 0.686 \\
D6 & 0.859 & 0.873 & 0.913 & 0.921 & 0.986 & 0.987 \\
D7 & 0.845 & 0.848 & 0.894 & 0.968 & 0.833 & 0.765 \\
D8 & 0.818 & 0.869 & 0.835 & 0.846 & 0.836 & 0.843 \\
D9 & 0.807 & 0.809 & 0.783 & 0.810 & 0.760 & 0.825 \\
\bottomrule
\end{tabular}
\begin{tablenotes}[flushleft]\footnotesize
\item \textit{Note:} Empirical coverage of the nominal 80\% interval $[q_{0.1}, q_{0.9}]$, averaged over lengths and replications; series-cluster bootstrap Monte Carlo SE at most 0.017.
\end{tablenotes}
\end{threeparttable}
\end{table}

\section{Response-surface regressions}
\label{si:response}

Tables~\ref{tab:si-response} and~\ref{tab:si-response2} report the regressions summarised in the section of the main text on
randomised parameters and departures from the assumptions.

\begin{table}[!htbp]
\centering
\begin{threeparttable}
\caption{Response-surface regressions of the robustness study, processes D1--D4.}
\label{tab:si-response}
\scriptsize
\begin{tabular}{llcccc}
\toprule
Process & Term & Coefficient & 95\% interval & $p$ & $R^2$ \\
\midrule
D1 & version: est period & 0.000 & [-0.002, 0.002] & 0.945 & 0.005 \\
D1 & version: heavy outliers & -0.032 & [-0.061, -0.003] & 0.029 & 0.005 \\
D1 & version: heavy tail & -0.022 & [-0.048, 0.004] & 0.091 & 0.005 \\
D1 & version: outliers & -0.010 & [-0.031, 0.010] & 0.334 & 0.005 \\
D1 & $\log_2 n$ & 0.005 & [-0.009, 0.020] & 0.490 & 0.005 \\
D1 & parameter phi & -0.018 & [-0.036, -0.000] & 0.045 & 0.005 \\
D1 & parameter sigma & 0.005 & [-0.011, 0.020] & 0.535 & 0.005 \\
D2 & version: est period & -0.000 & [-0.001, 0.001] & 0.795 & 0.013 \\
D2 & version: heavy outliers & -0.025 & [-0.052, 0.002] & 0.066 & 0.013 \\
D2 & version: heavy tail & -0.016 & [-0.041, 0.009] & 0.207 & 0.013 \\
D2 & version: outliers & 0.004 & [-0.014, 0.021] & 0.661 & 0.013 \\
D2 & $\log_2 n$ & 0.001 & [-0.015, 0.016] & 0.946 & 0.013 \\
D2 & parameter phi & -0.035 & [-0.055, -0.015] & 0.001 & 0.013 \\
D2 & parameter sigma & 0.011 & [-0.007, 0.029] & 0.224 & 0.013 \\
D2 & parameter theta & 0.003 & [-0.014, 0.020] & 0.726 & 0.013 \\
D3 & version: est period & 0.000 & [-0.004, 0.005] & 0.876 & 0.008 \\
D3 & version: heavy outliers & 0.026 & [-0.032, 0.084] & 0.381 & 0.008 \\
D3 & version: heavy tail & 0.020 & [-0.036, 0.075] & 0.487 & 0.008 \\
D3 & version: outliers & -0.004 & [-0.039, 0.032] & 0.843 & 0.008 \\
D3 & $\log_2 n$ & 0.017 & [-0.016, 0.050] & 0.306 & 0.008 \\
D3 & parameter phi & -0.055 & [-0.094, -0.016] & 0.006 & 0.008 \\
D3 & parameter sigma & 0.002 & [-0.035, 0.039] & 0.917 & 0.008 \\
D3 & parameter theta & 0.010 & [-0.027, 0.047] & 0.598 & 0.008 \\
D3 & parameter drift & -0.003 & [-0.040, 0.034] & 0.869 & 0.008 \\
D4 & version: est period & -0.004 & [-0.009, 0.001] & 0.141 & 0.233 \\
D4 & version: heavy outliers & -0.217 & [-0.261, -0.172] & 0.000 & 0.233 \\
D4 & version: heavy tail & 0.061 & [0.028, 0.095] & 0.000 & 0.233 \\
D4 & version: outliers & -0.219 & [-0.260, -0.178] & 0.000 & 0.233 \\
D4 & $\log_2 n$ & 0.238 & [0.210, 0.266] & 0.000 & 0.233 \\
D4 & parameter phi & -0.051 & [-0.084, -0.018] & 0.002 & 0.233 \\
D4 & parameter sigma & 0.056 & [0.022, 0.090] & 0.001 & 0.233 \\
D4 & parameter Phi & 0.098 & [0.064, 0.132] & 0.000 & 0.233 \\
D4 & parameter amp & -0.039 & [-0.071, -0.008] & 0.013 & 0.233 \\
\bottomrule
\end{tabular}
\begin{tablenotes}[flushleft]\footnotesize
\item \textit{Note:} One ordinary least-squares regression per process of the per-series $\log_2$ ratio of the MASE of TimesFM-3 to that of AutoARIMA on $\log_2 n$ (not standardised), the version (reference: random parameters with Gaussian innovations) and the drawn parameters, each standardised to mean 0 and standard deviation 1. Standard errors are clustered by parameter draw, because the five versions of a draw share its parameters and normal draws; 4\,000 series per process (3\,985 for D8, where series with a zero MASE denominator are excluded). Positive coefficients move the ratio against TimesFM-3.
\end{tablenotes}
\end{threeparttable}
\end{table}

\begin{table}[!htbp]
\centering
\begin{threeparttable}
\caption{Response-surface regressions of the robustness study, processes D5--D9.}
\label{tab:si-response2}
\scriptsize
\begin{tabular}{llcccc}
\toprule
Process & Term & Coefficient & 95\% interval & $p$ & $R^2$ \\
\midrule
D5 & version: est period & -0.086 & [-0.111, -0.061] & 0.000 & 0.073 \\
D5 & version: heavy outliers & -0.123 & [-0.175, -0.071] & 0.000 & 0.073 \\
D5 & version: heavy tail & -0.002 & [-0.039, 0.036] & 0.934 & 0.073 \\
D5 & version: outliers & -0.151 & [-0.191, -0.111] & 0.000 & 0.073 \\
D5 & $\log_2 n$ & 0.131 & [0.104, 0.158] & 0.000 & 0.073 \\
D5 & parameter sigma & 0.039 & [0.007, 0.072] & 0.018 & 0.073 \\
D5 & parameter amp & -0.035 & [-0.067, -0.003] & 0.030 & 0.073 \\
D5 & parameter alpha & -0.001 & [-0.034, 0.033] & 0.969 & 0.073 \\
D5 & parameter beta & -0.014 & [-0.048, 0.020] & 0.422 & 0.073 \\
D5 & parameter gamma & 0.028 & [-0.005, 0.061] & 0.100 & 0.073 \\
D5 & parameter trend & 0.019 & [-0.012, 0.051] & 0.231 & 0.073 \\
D6 & version: est period & -0.002 & [-0.004, 0.000] & 0.066 & 0.020 \\
D6 & version: heavy outliers & -0.026 & [-0.069, 0.018] & 0.245 & 0.020 \\
D6 & version: heavy tail & 0.018 & [-0.022, 0.058] & 0.384 & 0.020 \\
D6 & version: outliers & -0.023 & [-0.053, 0.007] & 0.140 & 0.020 \\
D6 & $\log_2 n$ & -0.046 & [-0.080, -0.013] & 0.006 & 0.020 \\
D6 & parameter sigma & -0.031 & [-0.071, 0.008] & 0.123 & 0.020 \\
D6 & parameter jump sd & -0.031 & [-0.065, 0.003] & 0.071 & 0.020 \\
D6 & parameter break pos & -0.039 & [-0.077, -0.001] & 0.043 & 0.020 \\
D6 & parameter sigma level & 0.033 & [-0.002, 0.068] & 0.064 & 0.020 \\
D6 & parameter jump sign & -0.019 & [-0.054, 0.017] & 0.295 & 0.020 \\
D7 & version: est period & -0.002 & [-0.011, 0.008] & 0.740 & 0.022 \\
D7 & version: heavy outliers & -0.074 & [-0.136, -0.013] & 0.017 & 0.022 \\
D7 & version: heavy tail & -0.031 & [-0.081, 0.018] & 0.216 & 0.022 \\
D7 & version: outliers & -0.045 & [-0.094, 0.005] & 0.075 & 0.022 \\
D7 & $\log_2 n$ & 0.031 & [-0.010, 0.072] & 0.138 & 0.022 \\
D7 & parameter sigma & -0.047 & [-0.098, 0.005] & 0.075 & 0.022 \\
D7 & parameter r & 0.071 & [0.020, 0.122] & 0.006 & 0.022 \\
D7 & parameter infl pos & -0.086 & [-0.140, -0.032] & 0.002 & 0.022 \\
D7 & parameter K & 0.010 & [-0.040, 0.059] & 0.703 & 0.022 \\
D8 & version: est period & -0.001 & [-0.003, 0.001] & 0.355 & 0.199 \\
D8 & version: heavy outliers & -0.093 & [-0.125, -0.061] & 0.000 & 0.199 \\
D8 & version: heavy tail & -0.047 & [-0.078, -0.016] & 0.003 & 0.199 \\
D8 & version: outliers & -0.025 & [-0.043, -0.007] & 0.007 & 0.199 \\
D8 & $\log_2 n$ & -0.099 & [-0.164, -0.034] & 0.003 & 0.199 \\
D8 & parameter p & 0.504 & [0.417, 0.591] & 0.000 & 0.199 \\
D8 & parameter lam & 0.012 & [-0.060, 0.084] & 0.748 & 0.199 \\
D9 & version: est period & -0.002 & [-0.003, -0.000] & 0.045 & 0.001 \\
D9 & version: heavy outliers & -0.019 & [-0.046, 0.008] & 0.164 & 0.001 \\
D9 & version: heavy tail & -0.007 & [-0.032, 0.018] & 0.592 & 0.001 \\
D9 & version: outliers & -0.013 & [-0.032, 0.007] & 0.196 & 0.001 \\
D9 & $\log_2 n$ & -0.002 & [-0.017, 0.013] & 0.807 & 0.001 \\
D9 & parameter phi & 0.001 & [-0.017, 0.020] & 0.894 & 0.001 \\
D9 & parameter a & -0.000 & [-0.019, 0.018] & 0.974 & 0.001 \\
D9 & parameter b & -0.004 & [-0.022, 0.014] & 0.646 & 0.001 \\
\bottomrule
\end{tabular}
\begin{tablenotes}[flushleft]\footnotesize
\item \textit{Note:} One ordinary least-squares regression per process of the per-series $\log_2$ ratio of the MASE of TimesFM-3 to that of AutoARIMA on $\log_2 n$ (not standardised), the version (reference: random parameters with Gaussian innovations) and the drawn parameters, each standardised to mean 0 and standard deviation 1. Standard errors are clustered by parameter draw, because the five versions of a draw share its parameters and normal draws; 4\,000 series per process (3\,985 for D8, where series with a zero MASE denominator are excluded). Positive coefficients move the ratio against TimesFM-3.
\end{tablenotes}
\end{threeparttable}
\end{table}

\section{Intermittent demand, parameter ranges and horizons}
\label{si:additional}

\tableref{tab:d8metrics} gives the error measures of the six pre-registered methods on the
intermittent process D8, and \tableref{tab:croston} compares TimesFM-3 with six Croston-family methods.
\tableref{tab:ranges} lists the parameter ranges of the randomised design, Figure~S1 shows accuracy by
forecast horizon, and Figure~S2 shows how the relative accuracy of TimesFM-3 and AutoARIMA varies with the
drawn parameters.

\begin{table}[!htbp]
\centering
\begin{threeparttable}
\caption{Accuracy on the intermittent process D8 under four error measures.}
\label{tab:d8metrics}
\small\setlength{\tabcolsep}{4pt}
\begin{tabular}{llrrrrrr}
\toprule
Metric & $n$ & Seas.\ naive & Theta & AutoETS & AutoARIMA & Combin. & \textbf{TimesFM-3} \\
\midrule
MASE & 24 & 1.019 & 0.999 & 1.047 & 0.996 & 1.010 & \textbf{0.780} \\
 & 48 & 1.066 & 0.984 & 0.988 & 0.962 & 0.974 & \textbf{0.708} \\
 & 96 & 0.932 & 0.923 & 0.926 & 0.924 & 0.923 & \textbf{0.683} \\
 & 200 & 1.005 & 0.943 & 0.939 & 0.939 & 0.940 & \textbf{0.694} \\
\addlinespace
RMSSE & 24 & 0.940 & 0.791 & 0.792 & \textbf{0.765} & 0.775 & 0.806 \\
 & 48 & 0.937 & 0.746 & 0.726 & \textbf{0.718} & 0.724 & 0.774 \\
 & 96 & 0.875 & 0.707 & \textbf{0.697} & 0.698 & 0.697 & 0.775 \\
 & 200 & 0.923 & 0.718 & \textbf{0.706} & 0.706 & 0.706 & 0.797 \\
\addlinespace
sMAPE & 24 & \textbf{86.696} & 172.741 & 169.988 & 165.908 & 170.489 & 191.774 \\
 & 48 & \textbf{84.690} & 172.522 & 170.746 & 170.622 & 171.026 & 196.183 \\
 & 96 & \textbf{84.171} & 170.801 & 169.574 & 170.065 & 169.949 & 197.610 \\
 & 200 & \textbf{85.990} & 170.003 & 169.189 & 169.315 & 169.236 & 198.556 \\
\addlinespace
Pinball & 24 & 0.634 & 0.395 & 0.402 & 0.380 & 0.388 & \textbf{0.325} \\
 & 48 & 0.653 & 0.381 & 0.368 & 0.360 & 0.365 & \textbf{0.307} \\
 & 96 & 0.614 & 0.348 & 0.341 & 0.341 & 0.341 & \textbf{0.288} \\
 & 200 & 0.628 & 0.354 & 0.344 & 0.345 & 0.345 & \textbf{0.291} \\
\bottomrule
\end{tabular}
\begin{tablenotes}[flushleft]\footnotesize
\item \textit{Note:} Means over $h = 1,\dots,12$ and 200 replications per length. The pinball loss is averaged over the nine deciles. Lower is better; the best value in each row is in bold.
\end{tablenotes}
\end{threeparttable}
\end{table}

\begin{table}[!htbp]
\centering
\begin{threeparttable}
\caption{Intermittent-demand baselines on the process D8 (post hoc analysis).}
\label{tab:croston}
\footnotesize\setlength{\tabcolsep}{3pt}
\begin{tabular}{llrrrrrrrr}
\toprule
Metric & $n$ & Croston & Croston-opt & Croston-SBA & ADIDA & IMAPA & TSB & AutoARIMA & \textbf{TimesFM-3} \\
\midrule
MASE & 24 & 1.042 & 1.024 & 1.025 & 0.972 & 0.972 & 0.971 & 0.996 & \textbf{0.780} \\
 & 48 & 0.990 & 0.992 & 0.974 & 0.947 & 0.954 & 0.966 & 0.962 & \textbf{0.708} \\
 & 96 & 0.935 & 0.948 & 0.921 & 0.926 & 0.923 & 0.916 & 0.924 & \textbf{0.683} \\
 & 200 & 0.947 & 0.948 & 0.934 & 0.941 & 0.942 & 0.948 & 0.939 & \textbf{0.694} \\
\addlinespace
RMSSE & 24 & 0.769 & 0.758 & 0.763 & \textbf{0.753} & 0.757 & 0.764 & 0.765 & 0.806 \\
 & 48 & 0.722 & 0.726 & 0.718 & \textbf{0.717} & 0.722 & 0.733 & 0.718 & 0.774 \\
 & 96 & 0.696 & 0.702 & \textbf{0.694} & 0.700 & 0.701 & 0.716 & 0.698 & 0.775 \\
 & 200 & 0.709 & 0.710 & 0.707 & 0.709 & 0.712 & 0.730 & \textbf{0.706} & 0.797 \\
\bottomrule
\end{tabular}
\begin{tablenotes}[flushleft]\footnotesize
\item \textit{Note:} Mean MASE and RMSSE over $h = 1,\dots,12$ and 200 replications per length. Croston-opt is Croston's method with optimised smoothing parameters and Croston-SBA the Syntetos--Boylan approximation. Lower is better; the best value in each row is in bold.
\end{tablenotes}
\end{threeparttable}
\end{table}

\begin{table}[!htbp]
\centering
\begin{threeparttable}
\caption{Parameter ranges of the randomised-parameter design.}
\label{tab:ranges}
\small\setlength{\tabcolsep}{4pt}
\begin{tabular}{ll}
\toprule
ID & Parameters, each drawn uniformly on the stated interval \\
\midrule
D1 & $\phi \in [0.10, 0.95]$, $\sigma \in [1, 5]$ \\
D2 & $\phi \in [0.10, 0.90]$, $\theta \in [-0.6, 0.8]$, $\sigma \in [1, 5]$ \\
D3 & $\phi \in [-0.5, 0.8]$, $\theta \in [-0.5, 0.6]$, drift $\in [0, 0.6]$, $\sigma \in [1, 5]$ \\
D4 & $\phi \in [0.1, 0.8]$, $\Phi \in [0.1, 0.8]$, seasonal amplitude $\in [5, 20]$, $\sigma \in [1, 5]$ \\
D5 & $\alpha \in [0.05, 0.5]$, $\beta \in [0.001, 0.03]$, $\gamma \in [0.05, 0.3]$, initial trend $\in [0, 0.5]$, amplitude $\in [5, 20]$, $\sigma \in [1, 5]$ \\
D6 & jump $\in [3, 15]\,\sigma$ with random sign, break at $[0.50, 0.85]\,n$, level noise $\in [0.2, 1]$, $\sigma \in [1, 5]$ \\
D7 & $r \in [0.04, 0.15]$, inflection at $[0.4, 0.8](n + H)$, $K \in [100, 250]$, $\sigma \in [1, 6]$ \\
D8 & demand probability $p \in [0.1, 0.6]$, size $1 + \mathrm{Pois}(\lambda)$ with $\lambda \in [1, 8]$ \\
D9 & $\phi \in [0.1, 0.9]$, $a \in [0.05, 0.15]$, $b \in [0.75, 0.90]$; if $a + b \geq 0.98$, $b$ is set to $0.97 - a$ \\
\bottomrule
\end{tabular}
\begin{tablenotes}[flushleft]\footnotesize
\item \textit{Note:} Parameters not listed are as in Table~2 of the main text. The intervals contain the
fixed values of the main design. The rule for D9 keeps the conditional variance stationary and puts
a small point mass at $b = 0.97 - a$. Seeds are disjoint from those of the main design; the five
versions share the parameter draws and, except for the heavy-tailed D8 sizes, the underlying normal
draws.
\end{tablenotes}
\end{threeparttable}
\end{table}

\begin{figure}[!htbp]
\centering
\includegraphics[width=\textwidth]{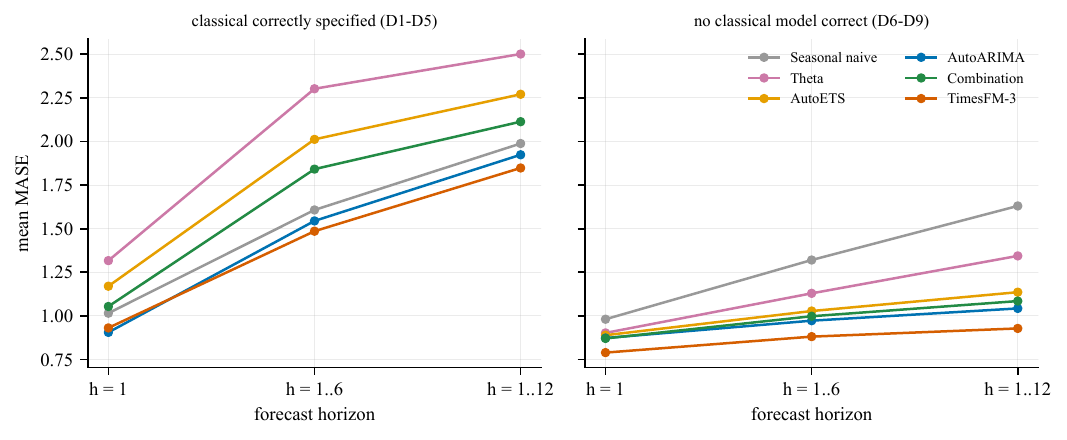}
\caption{Mean MASE by forecast horizon, averaged within each regime. Left: D1--D5, for which a classical
family is correctly specified. Right: D6--D9, for which none is, or only in the conditional mean (D9).}
\label{fig:si-horizon}
\end{figure}

\begin{figure}[!htbp]
\centering
\includegraphics[width=\textwidth]{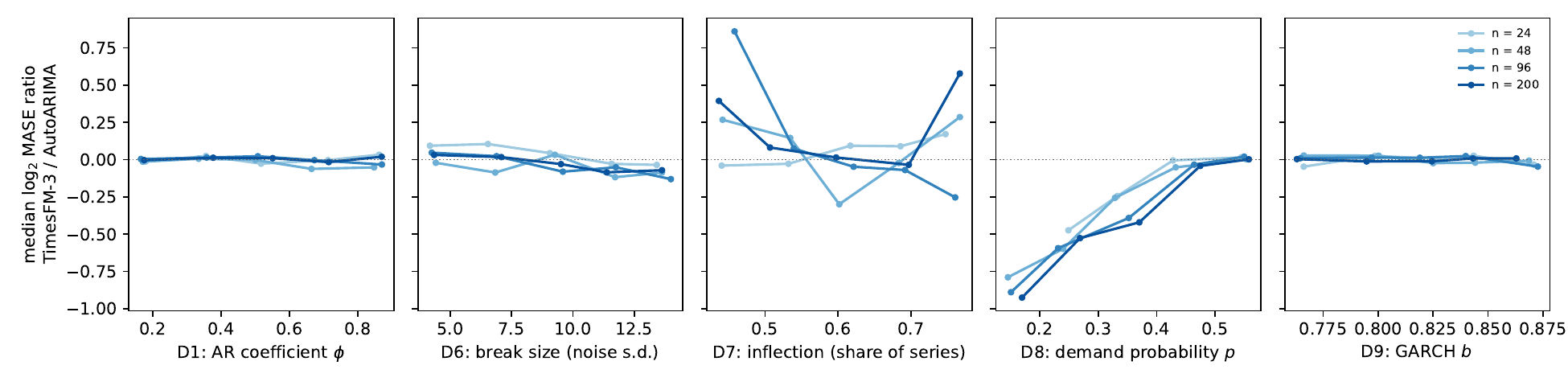}
\caption{Median $\log_2$ ratio of the MASE of TimesFM-3 to that of AutoARIMA, by quintile of the drawn
parameter, randomised-parameter design with Gaussian innovations. Values below zero favour TimesFM-3. For D8
the ratio compares median-type forecasts and is shown for completeness only.}
\label{fig:si-response}
\end{figure}

\section{Supplementary diagnostics}
\label{si:diagnostics}

\tableref{tab:si-worst} summarises each method's ratio to the best method of a scenario with and without the
two scenarios in which only two seasonal cycles are observed (D4 and D5 at $n = 24$), and \tableref{tab:extended}
gives the robustness summaries of the fifteen methods of Figure~5 of the main text, with bootstrap intervals.
\tableref{tab:si-forms}
lists the model forms chosen by AutoETS and AutoARIMA on the seasonal processes at $n = 24$ and $n = 48$,
and \tableref{tab:si-is} gives the 80\% interval score by process. Pooling all 180 tests of the
$h = 1$--12 slice into one Benjamini--Hochberg family gives 134 significant comparisons, 115 won by
TimesFM-3 (13--9 against AutoARIMA); pooling all 540 tests gives 132 and 113 (12--9), against 132 and 113
(11--9) with the pre-registered families.

\begin{table}[!htbp]
\centering
\begin{threeparttable}
\caption{Ratio of each method's mean MASE to the best in the same scenario, with and without the two-cycle scenarios.}
\label{tab:si-worst}
\small\setlength{\tabcolsep}{4pt}
\begin{tabular}{lcccccl}
\toprule
Method & Median & Geom.\ mean & 90th pct. & Worst & Worst excl. & Scenario of worst excl. \\
\midrule
Seasonal naive & 1.200 & 1.331 & 1.60 & 5.65 & 5.65 & D7, $n = 48$ \\
Theta & 1.251 & 1.458 & 2.97 & 4.69 & 4.69 & D4, $n = 200$ \\
AutoETS & 1.194 & 1.296 & 1.91 & 3.50 & 2.72 & D4, $n = 96$ \\
AutoARIMA & 1.032 & 1.118 & 1.37 & 2.22 & 1.62 & D7, $n = 96$ \\
Combination & 1.108 & 1.228 & 1.74 & 3.00 & 2.08 & D4, $n = 96$ \\
TimesFM-3 & 1.008 & 1.050 & 1.22 & 1.31 & 1.31 & D4, $n = 200$ \\
\bottomrule
\end{tabular}
\begin{tablenotes}[flushleft]\footnotesize
\item \textit{Note:} Main design, 36 scenarios, $h = 1, \dots, 12$. ``Excl.'' leaves out D4 and D5 at $n = 24$, where only two seasonal cycles are observed; the 90th percentile is taken over the 36 scenarios.
\end{tablenotes}
\end{threeparttable}
\end{table}

\begin{table}[!htbp]
\centering
\begin{threeparttable}
\caption{Robustness of fifteen methods across the 36 scenarios of the main design.}
\label{tab:extended}
\small\setlength{\tabcolsep}{4pt}
\begin{tabular}{lcccc}
\toprule
Method & Worst ratio to best & Mean ratio to best & Within 10\% (\%) & Scenarios won \\
\midrule
TimesFM-3, evaluator settings & 1.27 [1.26, 1.33] & 1.058 [1.055, 1.066] & 81 & 5 [2, 8] \\
TimesFM-3 & 1.31 [1.29, 1.36] & 1.063 [1.060, 1.072] & 78 & 0 [0, 4] \\
TimesFM-2.5 & 2.18 [2.04, 2.33] & 1.121 [1.116, 1.130] & 67 & 7 [2, 8] \\
TiRex & 2.24 [2.06, 2.45] & 1.121 [1.116, 1.131] & 64 & 0 [0, 4] \\
TimesFM-2.5, settings off & 2.26 [2.10, 2.44] & 1.124 [1.120, 1.134] & 67 & 3 [0, 5] \\
AutoARIMA & 2.22 [2.08, 2.39] & 1.131 [1.125, 1.142] & 64 & 8 [5, 10] \\
Chronos-2 & 2.75 [2.56, 2.97] & 1.139 [1.135, 1.150] & 67 & 4 [3, 6] \\
Combination (ETS, ARIMA, DOTM) & 2.99 [2.84, 3.16] & 1.241 [1.236, 1.253] & 47 & 1 [0, 2] \\
Combination (Theta, ETS, ARIMA) & 3.00 [2.85, 3.17] & 1.243 [1.237, 1.255] & 44 & 1 [1, 3] \\
Chronos-Bolt & 3.06 [2.95, 3.25] & 1.296 [1.291, 1.309] & 58 & 1 [0, 2] \\
AutoETS & 3.50 [3.28, 3.73] & 1.312 [1.304, 1.327] & 31 & 1 [0, 2] \\
Seasonal naive & 5.70 [5.47, 6.03] & 1.347 [1.337, 1.362] & 25 & 2 [1, 3] \\
DOTM & 4.66 [4.34, 4.99] & 1.413 [1.405, 1.429] & 22 & 1 [1, 2] \\
M4 Comb & 4.45 [4.18, 4.79] & 1.456 [1.447, 1.474] & 19 & 1 [0, 2] \\
Theta & 4.69 [4.37, 5.03] & 1.476 [1.467, 1.493] & 19 & 1 [1, 2] \\
\bottomrule
\end{tabular}
\begin{tablenotes}[flushleft]\footnotesize
\item \textit{Note:} Main design, 7\,200 series, mean MASE over $h = 1, \dots, 12$, ratios to the best of the fifteen methods in each of the 36 (process, length) scenarios. Brackets: 95\% intervals from 500 bootstrap resamples of the replications within each scenario. Mean ratio is the geometric mean over scenarios. Intervals for the original six methods from 2\,000 paired resamples are given in Section~5.1 of the main text. Scenarios won: the two TimesFM-3 configurations split the scenarios they win (among the original six methods TimesFM-3 wins 16), and ratios can differ slightly from Table~3 of the main text because the best method of a scenario is taken over fifteen methods (seasonal naive: 5.70 against 5.65).
\end{tablenotes}
\end{threeparttable}
\end{table}

\begin{table}[!htbp]
\centering
\begin{threeparttable}
\caption{Model forms chosen by AutoETS and AutoARIMA on the seasonal processes, given $m = 12$.}
\label{tab:si-forms}
\scriptsize\setlength{\tabcolsep}{4pt}
\begin{tabular}{llccp{4.3cm}p{4.3cm}}
\toprule
Process & $n$ & ETS seas.\ (\%) & ARIMA seas.\ (\%) & Most frequent ETS forms & Most frequent ARIMA orders \\
\midrule
D4 & 24 & 8.5 & 0.0 & ETS(A,N,N) (100), ETS(M,N,N) (83), ETS(A,N,A) (8) & (2,0,1)(0,0,0) (52), (3,0,1)(0,0,0) (44) \\
D4 & 48 & 98.0 & 100.0 & ETS(A,N,A) (67), ETS(M,N,M) (60), ETS(M,N,A) (42) & (1,0,0)(1,1,0) (52), (1,0,0)(0,1,1) (26) \\
D5 & 24 & 0.0 & 0.0 & ETS(A,N,N) (117), ETS(M,N,N) (83) & (2,0,1)(0,0,0) (72), (4,0,0)(0,0,0) (52) \\
D5 & 48 & 100.0 & 100.0 & ETS(M,N,M) (54), ETS(A,A,A) (40), ETS(M,A,A) (26) & (0,1,1)(0,1,1) (48), (0,0,0)(0,1,1) (35) \\
\bottomrule
\end{tabular}
\begin{tablenotes}[flushleft]\footnotesize
\item \textit{Note:} 200 replications per row, the series of the main design; counts out of 200 in parentheses. A form is seasonal when it has a seasonal component (ETS) or a non-zero seasonal order (ARIMA).
\end{tablenotes}
\end{threeparttable}
\end{table}

\begin{table}[!htbp]
\centering
\begin{threeparttable}
\caption{Scaled 80\% interval score by process, averaged over the four lengths.}
\label{tab:si-is}
\small\setlength{\tabcolsep}{4pt}
\begin{tabular}{lcccccc}
\toprule
Process & Seasonal naive & Theta & AutoETS & AutoARIMA & Combination & TimesFM-3 \\
\midrule
D1 & 8.49 & 7.80 & 7.67 & 6.36 & 6.97 & \textbf{6.05} \\
D2 & 9.05 & 9.28 & 9.09 & 6.98 & 8.00 & \textbf{6.86} \\
D3 & 23.33 & 23.31 & 22.12 & 22.03 & \textbf{21.01} & 21.24 \\
D4 & \textbf{4.58} & 16.77 & 11.04 & 5.65 & 9.65 & 5.69 \\
D5 & 4.64 & 7.22 & 6.51 & 4.29 & 5.59 & \textbf{3.98} \\
D6 & 9.35 & 7.13 & 7.21 & 7.53 & 7.28 & \textbf{4.55} \\
D7 & 14.14 & 8.59 & 5.91 & 5.17 & 5.51 & \textbf{4.55} \\
D8 & 10.11 & 4.60 & 4.50 & 4.40 & 4.47 & \textbf{3.54} \\
D9 & 8.56 & 7.22 & 7.07 & 5.82 & 6.44 & \textbf{5.67} \\
\bottomrule
\end{tabular}
\begin{tablenotes}[flushleft]\footnotesize
\item \textit{Note:} Interval score of $[q_{0.1}, q_{0.9}]$ with $\alpha = 0.2$ (width plus $2/\alpha$ times the distance by which the outcome falls outside the interval), averaged over the horizon and scaled by the MASE denominator, as for MSIS; lower is better, the best in bold. TimesFM-3 has the lowest score in 22 of the 36 (process, length) scenarios.
\end{tablenotes}
\end{threeparttable}
\end{table}

\begin{table}[!htbp]
\centering
\begin{threeparttable}
\caption{Classical methods in \pkg{forecast} (R) and \pkg{StatsForecast} (Python) on the seasonal processes: mean MASE over $h = 1, \dots, 12$.}
\label{tab:si-rcheck}
\footnotesize\setlength{\tabcolsep}{4pt}
\begin{tabular}{llccccccc}
\toprule
Process & $n$ & R ARIMA & SF ARIMA & R ETS & SF ETS & R Theta & SF Theta & R seas.\ (\%) \\
\midrule
D4 & 24 & 2.866 & 2.284 & 3.949 & 3.602 & 4.140 & 4.146 & 0 / 13 \\
D4 & 48 & 1.008 & 1.014 & 1.765 & 1.720 & 1.949 & 1.944 & 100 / 100 \\
D4 & 96 & 0.891 & 0.886 & 1.044 & 2.407 & 2.896 & 2.910 & 100 / 100 \\
D4 & 200 & 0.894 & 0.886 & 1.056 & 1.085 & 4.085 & 4.155 & 100 / 100 \\
D5 & 24 & 1.695 & 1.574 & 3.304 & 3.025 & 3.147 & 3.154 & 0 / 0 \\
D5 & 48 & 0.717 & 0.717 & 0.739 & 0.741 & 0.947 & 0.956 & 100 / 100 \\
D5 & 96 & 0.747 & 0.732 & 0.719 & 0.761 & 0.759 & 0.757 & 100 / 100 \\
D5 & 200 & 0.631 & 0.626 & 0.644 & 0.681 & 0.760 & 0.759 & 100 / 100 \\
\bottomrule
\end{tabular}
\begin{tablenotes}[flushleft]\footnotesize
\item \textit{Note:} The same 200 series per row. R: \code{auto.arima}, \code{ets} and \code{thetaf} of \pkg{forecast} 9.0.2 at their defaults, with frequency 12; SF: \pkg{StatsForecast} 2.1.1 as in the main text. R seas.: share of series given a seasonal form by \code{auto.arima} / \code{ets}. No R fit failed. On D4 at $n = 96$ \pkg{StatsForecast}'s AutoETS settles on a near-zero seasonal smoothing parameter with a much lower likelihood than R's fit of the same form.
\end{tablenotes}
\end{threeparttable}
\end{table}

\section{A longer horizon}
\label{si:h48}

The main design forecasts 12 steps. To see whether its conclusions depend on that choice, the nine
processes were regenerated with the same seeds at length $n + 48$ and forecast 48 steps ahead by the six
pre-registered methods, TimesFM-2.5 and Chronos-2 (post hoc). Because the inflection of D7 sits at
$0.6(n + H)$, the longer horizon moves it into the context for $n = 200$: the series has reached its plateau
before the forecast origin. \tableref{tab:si-h48} and Figure~S3 summarise the results.

\begin{table}[!htbp]
\centering
\begin{threeparttable}
\caption{Longer horizon ($H = 48$): worst ratio to the scenario-best and 80\% coverage by block of steps.}
\label{tab:si-h48}
\small\setlength{\tabcolsep}{4pt}
\begin{tabular}{lccccccc}
\toprule
Method & \multicolumn{3}{c}{Worst ratio, D7 excluded} & \multicolumn{3}{c}{Coverage 80\%} & D7, $n = 200$ \\
 & 1--12 & 13--24 & 25--48 & 1--12 & 13--24 & 25--48 & MASE 1--48 \\
\midrule
TimesFM-3 & 1.31 & 1.39 & 1.80 & 0.788 & 0.774 & 0.760 & 8.37 \\
TimesFM-2.5 & 1.42 & 1.59 & 1.77 & 0.781 & 0.734 & 0.707 & 12.22 \\
Chronos-2 & 2.94 & 2.71 & 3.33 & 0.787 & 0.762 & 0.761 & 5.30 \\
AutoARIMA & 2.11 & 1.60 & 1.41 & 0.760 & 0.739 & 0.720 & 2.51 \\
AutoETS & 3.49 & 2.17 & 1.67 & 0.779 & 0.790 & 0.784 & 2.35 \\
Theta & 4.65 & 3.66 & 6.27 & 0.731 & 0.757 & 0.768 & 2.73 \\
Combination & 2.92 & 2.00 & 2.17 & 0.788 & 0.812 & 0.814 & 2.22 \\
Seasonal naive & 1.66 & 1.88 & 1.87 & 0.805 & 0.780 & 0.753 & 1.11 \\
\bottomrule
\end{tabular}
\begin{tablenotes}[flushleft]\footnotesize
\item \textit{Note:} The nine processes regenerated with the same seeds at length $n + 48$, 200 replications per scenario; ratios to the best of the eight methods in each of the 32 (process, length) scenarios without D7. Coverage is averaged over all 36 scenarios. The last column is the mean MASE over the 48 steps on D7 at $n = 200$, where the process has reached its plateau before the forecast origin.
\end{tablenotes}
\end{threeparttable}
\end{table}

\begin{figure}[!htbp]
\centering
\includegraphics[width=\textwidth]{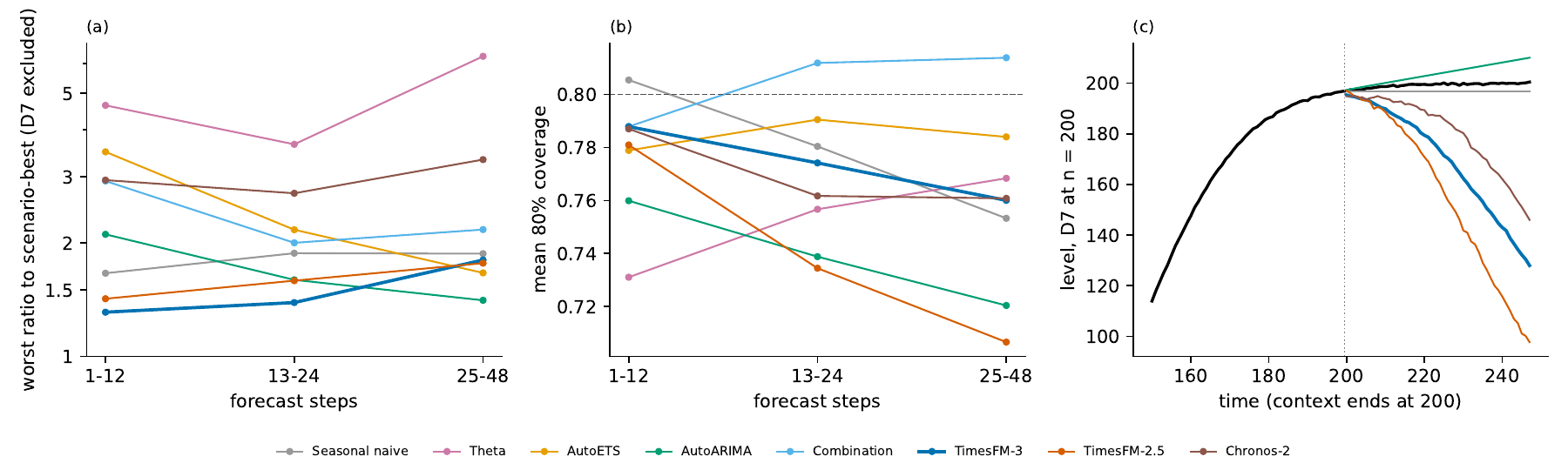}
\caption{Longer horizon ($H = 48$). (a) Worst ratio to the scenario-best method by block of forecast steps, D7
excluded. (b) Mean coverage of the nominal 80\% intervals by block; the dashed line is the nominal level.
(c) D7 at $n = 200$: mean of the 200 series (black) and mean point forecasts over the 48 steps.}
\label{fig:si-h48}
\end{figure}

\section{Further checks}
\label{si:further}

\tableref{tab:si-count} compares count-data benchmarks with the history benchmark and TimesFM-3 on the
intermittent process, \tableref{tab:si-postcutoff} reports the tier of data observed after the documented
training corpora, and \tableref{tab:si-simult} counts the scenarios separated by simultaneous and by
per-scenario intervals.

\begin{table}[!htbp]
\centering
\begin{threeparttable}
\caption{Count-data benchmarks on the intermittent process D8.}
\label{tab:si-count}
\footnotesize\setlength{\tabcolsep}{4pt}
\begin{tabular}{lcccccccc}
\toprule
& \multicolumn{4}{c}{Scaled pinball loss} & \multicolumn{4}{c}{RMSSE of the mean} \\
Method & 24 & 48 & 96 & 200 & 24 & 48 & 96 & 200 \\
\midrule
iETS & 0.306 & 0.298 & 0.283 & 0.287 & 0.737 & 0.712 & 0.693 & 0.704 \\
Negative binomial & 0.320 & 0.304 & 0.289 & 0.295 & 0.744 & 0.710 & 0.692 & 0.703 \\
TSB compound & 0.330 & 0.314 & 0.298 & 0.306 & 0.764 & 0.732 & 0.714 & 0.729 \\
History: deciles; mean & 0.314 & 0.298 & 0.282 & 0.286 & 0.744 & 0.710 & 0.692 & 0.703 \\
TimesFM-3 (mean of deciles) & 0.325 & 0.307 & 0.288 & 0.291 & 0.760 & 0.717 & 0.694 & 0.705 \\
\bottomrule
\end{tabular}
\begin{tablenotes}[flushleft]\footnotesize
\item \textit{Note:} 200 replications per length (iETS failed on 3 series at $n = 24$). iETS: \code{smooth::adam(y, "MNN", occurrence = "auto")} with simulated quantiles; negative binomial fitted by maximum likelihood to the context; TSB compound: TSB occurrence probability ($\alpha = 0.2$) times the empirical distribution of the non-zero sizes. Point forecasts are the predictive means.
\end{tablenotes}
\end{threeparttable}
\end{table}

\begin{table}[!htbp]
\centering
\begin{threeparttable}
\caption{Data observed after the documented training corpora: 101 FRED-MD series, January 2025 to June 2026.}
\label{tab:si-postcutoff}
\small\setlength{\tabcolsep}{4pt}
\begin{tabular}{lccccc}
\toprule
Method & Mean MASE & Median MASE & Mean MASE, 2025-11 on & Coverage 80\% & SPL \\
\midrule
AutoARIMA & 0.540 & 0.411 & 0.722 & 0.913 & 0.241 \\
TiRex & 0.556 & 0.439 & 0.746 & 0.864 & 0.229 \\
Combination & 0.556 & 0.445 & 0.756 & 0.880 & 0.250 \\
AutoETS & 0.572 & 0.455 & 0.774 & 0.908 & 0.262 \\
Chronos-Bolt & 0.582 & 0.505 & 0.776 & 0.887 & 0.238 \\
Chronos-2 & 0.593 & 0.424 & 0.808 & 0.798 & 0.245 \\
Theta & 0.602 & 0.479 & 0.833 & 0.843 & 0.270 \\
TimesFM-3 & 0.607 & 0.426 & 0.857 & 0.838 & 0.242 \\
TimesFM-2.5 & 0.608 & 0.404 & 0.833 & 0.850 & 0.254 \\
Seasonal naive & 0.896 & 0.664 & 1.145 & 0.911 & 0.381 \\
\bottomrule
\end{tabular}
\begin{tablenotes}[flushleft]\footnotesize
\item \textit{Note:} FRED-MD, August 2026 vintage, raw levels; context January 2005 to December 2024, horizon 18. MASE scaled by the in-sample seasonal-naive error ($m = 12$). The later window (November 2025 to June 2026) post-dates the release of every model evaluated.
\end{tablenotes}
\end{threeparttable}
\end{table}

\begin{table}[!htbp]
\centering
\begin{threeparttable}
\caption{Scenarios in which TimesFM-3 is separated from each opponent: simultaneous against per-scenario intervals.}
\label{tab:si-simult}
\small\setlength{\tabcolsep}{4pt}
\begin{tabular}{lcccc}
\toprule
& \multicolumn{2}{c}{Simultaneous (max-$t$)} & \multicolumn{2}{c}{Per scenario} \\
Opponent & TimesFM-3 better & Opponent better & TimesFM-3 better & Opponent better \\
\midrule
AutoARIMA & 8 & 4 & 11 & 9 \\
Seasonal naive & 17 & 7 & 22 & 8 \\
Theta & 23 & 0 & 28 & 1 \\
AutoETS & 17 & 0 & 25 & 3 \\
Combination & 15 & 2 & 27 & 3 \\
\bottomrule
\end{tabular}
\begin{tablenotes}[flushleft]\footnotesize
\item \textit{Note:} Median paired $\log_2$ ratio of MASE over $h = 1, \dots, 12$ in each of the 36 scenarios; 95\% intervals from 2\,000 bootstrap resamples of the replications, simultaneous over the 36 scenarios (studentised maximum) or per scenario (percentile).
\end{tablenotes}
\end{threeparttable}
\end{table}

\end{document}